\documentclass[journal=jacsat,manuscript=article]{achemso}

\usepackage[version=3]{mhchem} 
\usepackage{xcolor} 

\author{Pieter J. van Essen}
\email{p.vessen@arcnl.nl}
\author{Zhonghui Nie}
\author{Brian de Keijzer}
\author{Peter M. Kraus}
\affiliation[ARCNL]
{Advanced Research Center for Nanolithography, Science Park 106, 1098 XG, Amsterdam, The Netherlands}
\altaffiliation{Department of Physics and Astronomy, and LaserLaB, Vrije Universiteit, De Boelelaan 1105,1081 HV Amsterdam, The Netherlands}
\email{p.kraus@arcnl.nl}
\title[]
  {Towards complete all-optical emission control of high-harmonic generation from solids}

\abbreviations{}

\begin{document}

\begin{tocentry}
\begin{figure}[H]
    \centering
    \includegraphics[width=\textwidth]{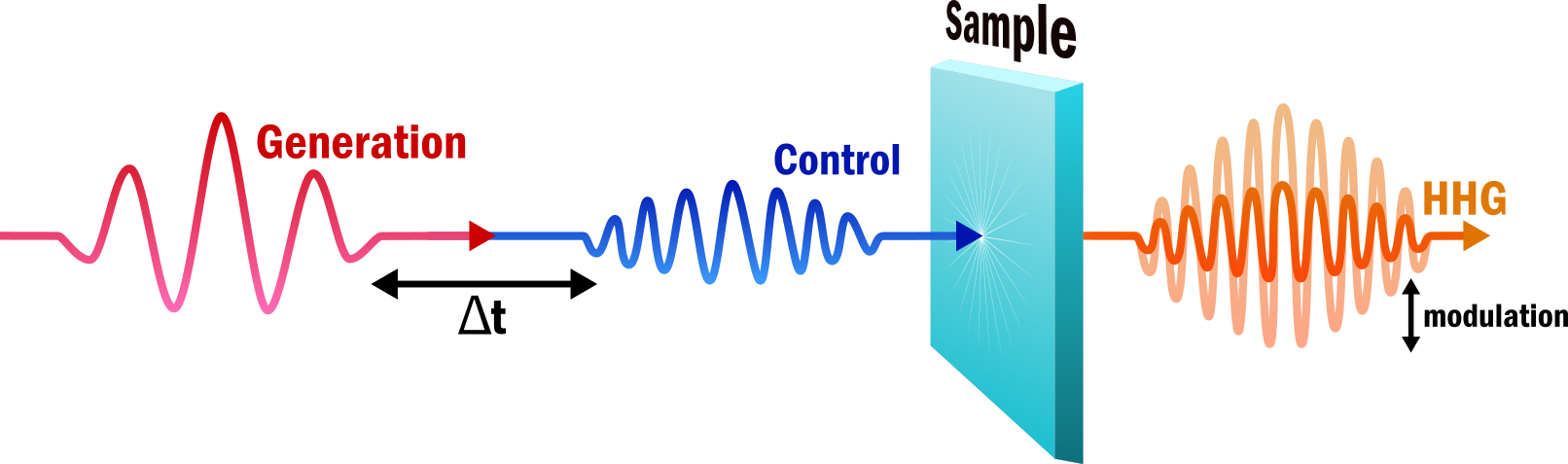}
    \label{fig:enter-label}
\end{figure}
\end{tocentry}

\begin{abstract}
Optical modulation of high-harmonics generation in solids enables the detection of material properties such as the band structure and promising new applications such as super-resolution imaging in semiconductors. Various recent studies have shown optical modulation of high-harmonics generation in solids, in particular, suppression of high-harmonics generation has been observed by synchronized or delayed multi-pulse sequences. Here we provide an overview of the underlying mechanisms attributed to this suppression and provide a perspective on the challenges and opportunities regarding these mechanisms. All-optical control of high-harmonic generation allows for femtosecond, and in the future possibly subfemtosecond, switching, which has numerous possible applications: These range from super-resolution microscopy, to nanoscale controlled chemistry, and highly tunable nonlinear light sources.
\end{abstract}

\newpage
\section{Introduction}
High-harmonic generation (HHG) allows for the coherent generation of ultra-short pulses which has enabled the field of attosecond science for the study of ultra-fast phenomena \cite{2007Corkum,Calegari2016,Li2020,Vampa2017,kraus18a,kraus18b,woerner17a}, nanoscale microscopy via coherent diffraction imaging \cite{jansen18a,loetgering2021tailoring}, and increasingly applications within the semiconductor industry for wafer metrology \cite{porter23a}.

Originally done with gases, recently there has been an increasing interest in HHG from solids \cite{Vampa2017,Goulielmakis2022,Huttner2016,Ghimire2011}. Solids are particularly interesting with regards to HHG as the generation process depends strongly on the electron dynamics which can vary widely between solids. Various different solids have been used for HHG, including but not limited to semiconductors: Si\cite{Vampa2016}, SiO \cite{Luu2015}, MgO \cite{You2017}, and ZnO \cite{Ghimire2011,Wang2017,Xu2022,Wang2023}, monolayers: graphene \cite{Cheng2020,Yoshikawa2017}, MoS$_2$ \cite{Wang2022,Heide2022}, and WSe$_2$ \cite{Nagai2023}, strongly correlated electron materials: VO$_2$ \cite{Bionta2021} and NbO$_2$ \cite{nie23a}, perovskites: MAPbBr$_3$ \cite{Geest23a} and even metals: TiN \cite{Korobenko2021}. While this article focuses on HHG from solids, high-harmonics have also been generated from liquids \cite{heissler14a,luu18b} and plasmas \cite{mathijssen23a,ganeev07a}. This wide range of materials indicates the generality of the HHG process. \\
Lessons from research on atomic and molecular gas-phase HHG teach us that HHG can be controlled to great extents by adding additional laser pulses. Important examples include the emission control of odd and even harmonic orders in multi-color laser fields \cite{dudovich06a,RoscamAbbing2020,roscam21a}, controlling the harmonic emission intensity by aligning \cite{sakai03a,rupenyan13a} and orienting molecules \cite{frumker12a,kraus12c,kraus14b,kraus15a}, as well as controlling emission by photoexciting molecules \cite{woerner10b,kraus12a,ruf12a,kraus12b,kraus13b,baykusheva14a,zhang15b}.\\
New developments in high-harmonic generation in solids provide an exciting prospect in enabling the study of microscopic electron dynamics \cite{Heide2022,Bionta2021,Hohenleutner2015}, highly tunable (EUV) light sources \cite{roscam22a,korobenko22a,franz19a}, ultra-fast all-optical signal modulation \cite{Xu2022,Wang2023,Cheng2020}, and super-resolution imaging \cite{Murzyn2023}. Instrumental for all these advancements is controlling high-harmonic generation. The literature on controlling gas HHG strongly suggests that all-optical emission control in solids is feasible. Recent works \cite{Wang2017,Xu2022,Wang2023,Wang2022,Heide2022,Nagai2023,Bionta2021,Cheng2020,Nishidome2020} have shown strong modulation of high-harmonic generation, most notably strong suppression was observed. While the consistent observation of HHG suppression might be considered contrary to the dream of achieving extremely efficient solid HHG sources, tunable suppression does allow for advancements in applications such as those mentioned above. The power of suppression is very well exemplified in the recent demonstration of label-free super-resolution imaging in solids using harmonic deactivation microscopy (HADES) \cite{Murzyn2023}, which has links to the successful fluorescence-based super-resolution technique stimulated emission depletion (STED). Here, the most crucial part is complete signal suppression and the ability to saturate this suppression. Additionally, the understanding of suppression of high-harmonics may very well be the key to increasing high-harmonic generation for the purpose of an all-solid HHG source. Thus, for the development and optimization of high-harmonic modulation, it is important to obtain a more complete understanding of the underlying principles. In this Perspective, we will discuss the recent developments in the modulation of high-harmonic from solids, provide an overview of the mechanism used to explain these recent observations, and provide an outlook into their main challenges.  \\
As we aim to provide a general overview, we refrain from going into detail discussing any material-specific properties. For the most part, our discussion will assume conventional semiconductor materials with the exception of the section on phase transitions where we will discuss strongly correlated electron materials. \\

\section{High-harmonic generation in solids}
\begin{figure}[H]
    \centering
    \includegraphics[width=\textwidth]{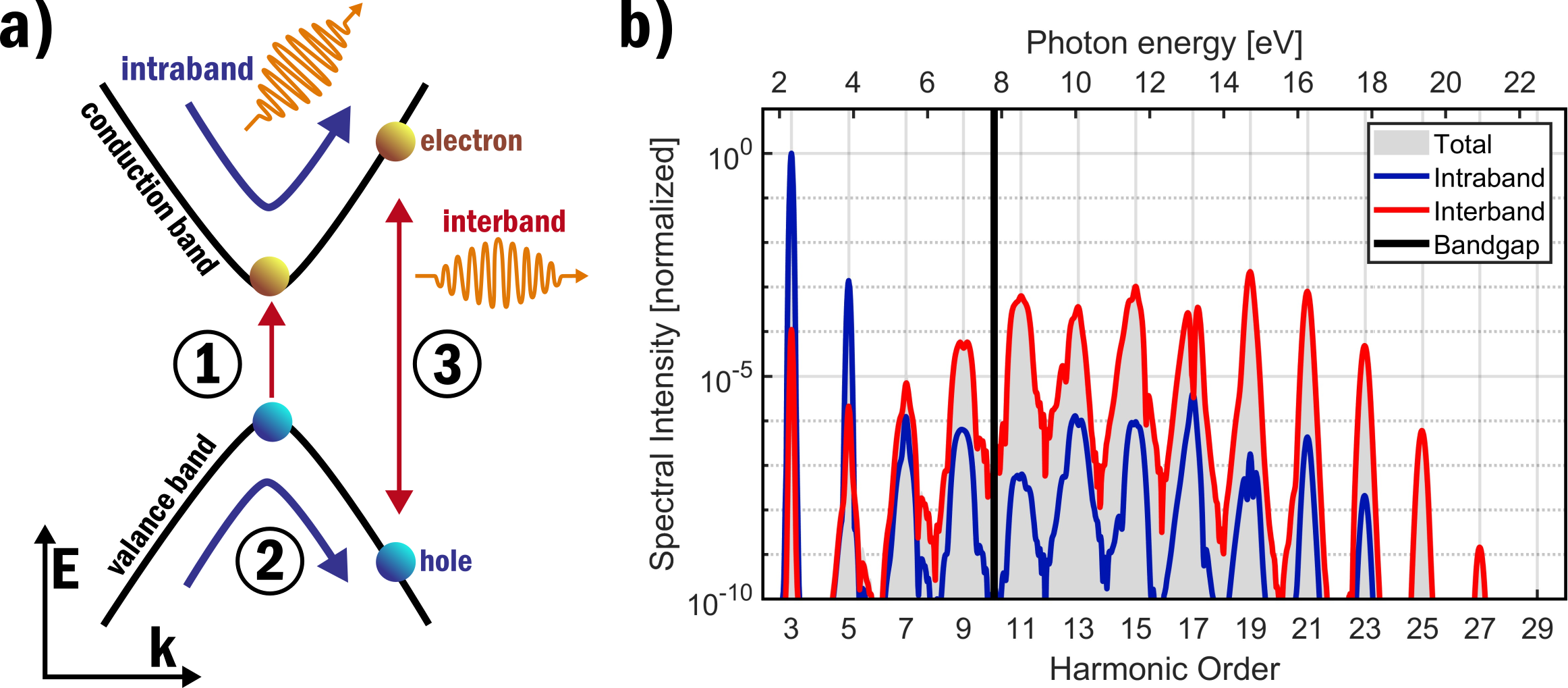}
    \caption{In (a) a schematic of the three-step model for high-harmonic generation is shown in reciprocal space for a solid. In step \textbf{1}, the strong external electric field excites an electron-hole pair. In step \textbf{2}, after excitation, the electron-hole pair is accelerated by the strong electric field gaining kinetic energy. Acceleration along the non-parabolic bands of a solid generates the intraband current. In step \textbf{3}, the electron-hole pair recombines which generates the interband current. The total high-harmonic signal is the combination of the intraband and interband contributions. In (b) an HHG spectrum is shown which is simulated using a one-dimensional SBE simulation with two sinusoidal bands with a 7.8 eV bandgap and a 1 TW/cm$^2$ generation pulse at 1600 nm. In blue (red) the intraband (interband) contributions to the HHG signal are shown.}
    \label{fig:3steps}
\end{figure}

High-harmonic generation is a highly non-linear strong-field effect. Harmonics are generated from a high-intensity fundamental driver, where the harmonic frequencies $\omega_n$ are multiples of the fundamental frequency $\omega_0$:
\begin{equation}
    \omega_n = n\cdot\omega_0.
\end{equation}
Normally only odd harmonics are generated due to the conservation of inversion symmetry \cite{Goulielmakis2022}, however, it is possible to also generate even harmonics by breaking this inversion symmetry either by modulation of the driving field \cite{Vampa2016,Luu2018} or by using specific materials \cite{Jia2020,Hohenleutner2015,Liu2017}. In solids intensities around GW/cm$^2$ to tens of TW/cm$^2$ are used for HHG, which is reached by the use of femtosecond pulsed lasers. \\
HHG is a strong-field effect where the electric field strength of the fundamental is in the same order of magnitude as the effective Coulomb force experienced by the valence electrons \cite{Ghimire2011}. As a result, HHG depends strongly on the fundamental electric field as well as the material itself. This also means that an accurate description of HHG in solids requires both a detailed description of the fundamental field and of the electronic structure of the material. 
For lower-intensity $n$-photon processes an intensity scaling of $I_n\propto I_0^n$ is
observed, which matches the predictions of perturbation theory \cite{Ward1965}. For higher harmonics clear deviations for these ideal scalings are observed \cite{Ghimire2011,Goulielmakis2022}. This indicates that HHG can not simply be described using perturbation theory.\\
The microscopic process underpinning high-harmonic generation is ultra-fast and takes place on the attosecond timescale within an optical cycle of the fundamental. The harmonics are generated via the induced movement of charge carriers by the strong electric field. This effective current can be separated into the intraband and interband currents which are respectively related to transitions to different momenta within the same band and between different electronic bands respectively. Solid high-harmonic generation can intuitively be understood in terms of a three-step model which is an adapted version of the three-step model used to explain gas high-harmonic generation \cite{Vampa2017}. In Fig. \ref{fig:3steps} the three-step model for solids is schematically shown. 
In the first step of the three-step model, the strong generation field excites an electron from a valence band to a conduction band, at the same time a hole is generated in the valance band. This excitation is enabled by the strong-field deforming the Coulomb potential to such an extent that tunneling becomes possible. In the second step, the excited electron and hole are accelerated by the strong electric field within their respective band gaining kinetic energy. During the acceleration of the electron-hole pair, the varying effective mass along the bands allows for the generation of the intraband current. Notably, this current is not found in gas HHG as the effective mass of charge carriers in free space is constant. As the direction of the electric field changes within an optical cycle, the trajectory of the electron-hole pair can be such that they recombine which is the third and final step. The recombination of the electron-hole pairs generates the interband current.\\
The significance of the intraband and interband contributions varies between harmonic orders. The intuitive distinction can be made that the intraband current is more important for the lower-order below-bandgap harmonics. In contrast, the interband current is more important for the higher-order above-bandgap harmonics, this distinction can also be seen in the spectrum shown in Fig. \ref{fig:3steps}b. This separation can classically be understood by considering that the energy differences within a band are small compared to the bandgap. \\
The semi-classical description as given above does have its limits and to really obtain a predictive description of HHG in solids the electron dynamics have to be considered quantum mechanically. This semi-classical model does however provide a very effective framework in which to understand the HHG process and the underlying suppression mechanisms.\\
A more quantitative description of the system which accounts for the quantum mechanical nature can be obtained using the semiconductor Bloch equations (SBE) \cite{Lindeberg1998,Haug2004,Yue2021}. Using the length gauge in the dipole approximation and the Bloch basis to describe the electronic states the equation of motion (EOM) that is obtained for the density matrix elements with momentum $\mathbf{k}$ is given by \cite{Yue2021}:
\begin{equation}
    \begin{split}
        i \frac{\partial}{\partial t}\rho^\mathbf{k}_{mn}(t) =&
    \left(\epsilon^\mathbf{k}_{m}-\epsilon^\mathbf{k}_{n}\right)\rho^\mathbf{k}_{mn}(t)
    - \mathbf{F}(t) \cdot \sum_l\left[\mathbf{d}^\mathbf{k}_{ml}\rho^\mathbf{k}_{ln}(t) - \mathbf{d}^\mathbf{k}_{ln}\rho^\mathbf{k}_{ml}(t)\right] \\
    &
    + i \mathbf{F}(t)\cdot \nabla_\mathbf{k}\rho^\mathbf{k}_{mn}(t) -i\frac{1-\delta_{mn}}{T_2}\rho_{mn}^\mathbf{k}(t),        
    \end{split}
    \label{eq:EOM}
\end{equation}
$\mathbf{F}(t)$ here indicates the electric field, $\epsilon_n^{\mathbf{k}}$ is the energy level of the state, $\mathbf{d}_{nm}^\mathbf{k}$ is the dipole coupling between states $n$ and $m$, $T_2$ is the dephasing time, and $\delta_{mn}$ is the Dirac function. The diagonal elements of the density matrix ($m=n$) are the populations in the bands while the off-diagonal elements ($m\neq n$) are the coherences between the bands. 
The right-hand side of the EOM contains four distinct terms. The first term is a phase term and is dependent on the energy difference between the bands, the second term depends on the dipole couplings and describes transitions between the different bands, the third term depends on the derivative in k-space and describes the movement of carriers within a band, and the fourth term is a phenomenological damping term which describes the dephasing between the carriers. Dephasing will be discussed in more detail in the section on excitation-induced dephasing.
The interband and intraband current can be expressed as:
\begin{equation}
        \begin{split}
    \mathbf{j}_{\text{intra}}(t) & = \sum_{\mathbf{k}\in \text{BZ}}\sum_{m} \nabla_\mathbf{k}\epsilon^\mathbf{k}_{m}\rho^\mathbf{k}_{mm}(t),
    \\
    \mathbf{j}_{\text{inter}}(t) & = \sum_{\mathbf{k}\in \text{BZ}}\frac{\partial}{\partial t}\sum_{m\neq n}\mathbf{d}^\mathbf{k}_{mn}\rho^\mathbf{k}_{nm}(t).
    \\ 
    \end{split}
\end{equation}
BZ here refers to the Brillouin zone. We see that the intraband current is due to the movement of the population through a band, while the interband current is due to the time derivative of the coherent dipole coupling between bands. 
Important to note is that the intraband and interband currents are not entirely independent as the EOM couples the population and coherence via the dipole coupling. The spectral intensity can be obtained by applying a Fourier transform to the total current:
\begin{equation}
    S(\omega) = |\mathcal{F}\{\mathbf{j}_{\text{intra}}(t)+\mathbf{j}_{\text{inter}}(t)\}|^2.
\end{equation}

\section{Time-resolved high-harmonic generation}
\begin{figure}[H]
    \centering
    \includegraphics[width=\textwidth]{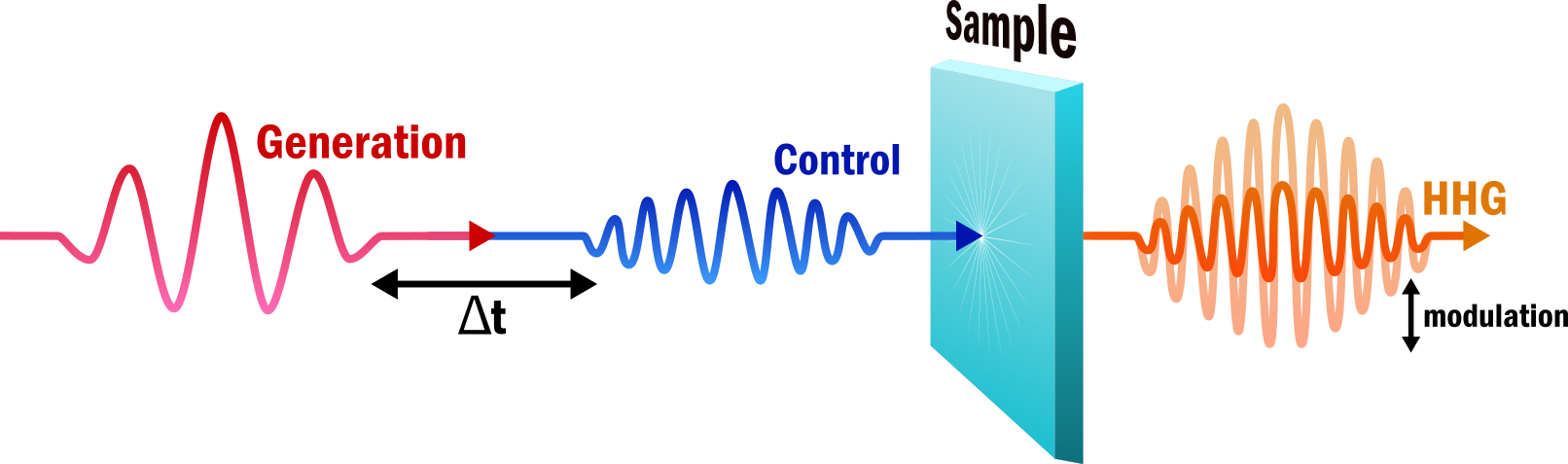}
    \caption{Schematic of the pump-probe system used to perform time-resolved high-harmonic generation. A high-intensity generation pulse is used to generate high-harmonics from a sample. An in space and time closely spaced control pulse is subsequently used to modulate the HHG. In the majority of measurements, the control pulse has a wavelength shorter than that of the generation pulse. By varying the delay time $\Delta t$ between the generation and control pulse time-resolved measurements are obtained. For positive delay times, the control pulse arrives before the generation pulse while for negative delay times, the control pulse arrives after the generation pulse.}
    \label{fig:pumpprobe_general}
\end{figure}

\begin{figure}[H]
    \centering
    \includegraphics[width=\textwidth]{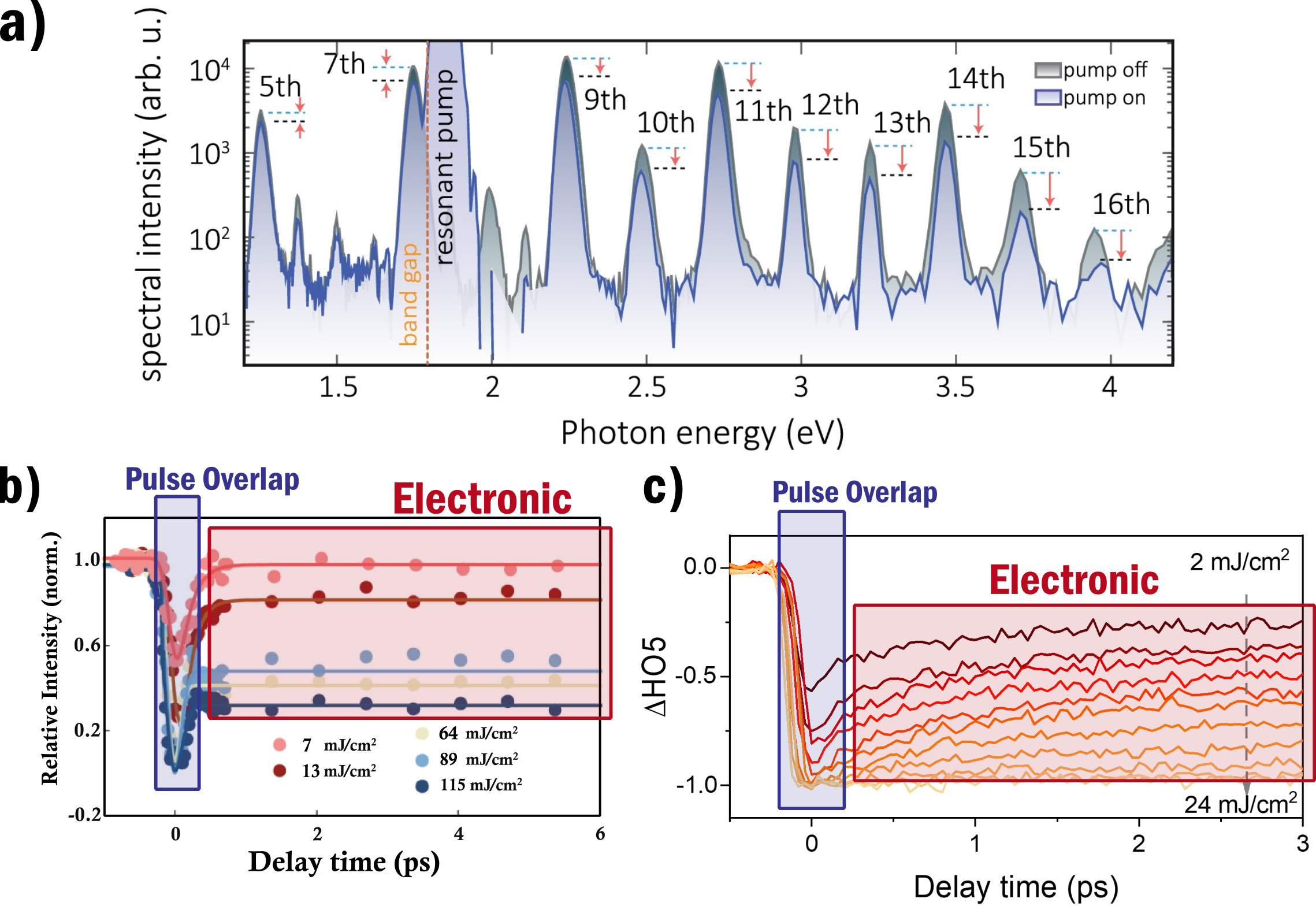}
    \caption{
    In (a) the harmonic spectra by Ref. \cite{Heide2022} from MoS$_2$ using 5000 nm as generation wavelength and 660 nm as control wavelength. The two different spectra show the control pulse being respectively on and off, the delay time is 1 ps. The arrows indicate the decrease in intensity for each of the harmonics. 
    In (b) the suppression curve measured by Ref. \cite{Xu2022} is shown for the fifth harmonic from ZnO using 2350 nm and 800 nm respectively as generation and control wavelengths.
    In (c) the suppression curve measured by Ref. \cite{nie23a} is shown for the fifth harmonic from NbO$_2$ using 2000 nm and 400 nm respectively as generation and control wavelengths. We note that the curves in (b) and (c) show very similar behaviour although they are measured from vastly different materials. A sharp drop in HHG yield is observed when the generation and control pulse overlap, while for longer delay times we see a gradual recovery of the intensity. In both (b) and (c) the overlap and electronic recovery region are indicated explicitly. Increased suppression is found for increasing control intensity.  }
    \label{fig:Suppression}
\end{figure}

Modulation of high-harmonic generation is achieved by introducing a control pulse, enabling pump-probe style measurements as shown in Fig. \ref{fig:pumpprobe_general}. The control pulse is made to spatially overlap the generation pulse such that it can affect the HHG process. By varying the delay time between the generation and control pulses time-resolved spectroscopy can be performed. Positive time delays refer to the situation where the control pulse arrives before the generation pulse while negative time delays refer to the situation where the generation pulse arrives first. The intensity used for the control pulse varies widely but is often lower than that of the generation pulse so as to not generate harmonics itself or damage the sample.
Different implementations of this same measurement scheme have been used in nearly all recent works which have shown modulation of HHG in solids \cite{Wang2017,Xu2022,Wang2023,Wang2022,Heide2022,Nagai2023,Bionta2021,Cheng2020}. An overview of these recent works with the materials and wavelengths used can be found in table \ref{table:overview}.\\ 
Although various materials have been studied using significantly different measurement conditions the results have shown some rather noticeable consistencies, the most important of these being strong suppression. Figure \ref{fig:Suppression}a shows the HHG spectra measured by Ref. \cite{Heide2022} from ZnO for a 1 ps pulse delay where increasing suppression is observed for the harmonic orders. In Fig. \ref{fig:Suppression}b the suppression of the fifth harmonic from ZnO measured by Ref. \cite{Xu2022} is shown as a function of delay time. Per comparison, Fig. \ref{fig:Suppression}c shows a similar figure but now for the fifth harmonic measured from NbO$_2$ by Ref. \cite{nie23a}. The suppression curves shown in Fig. \ref{fig:Suppression}b and \ref{fig:Suppression}c are exemplars for HHG suppression observed in solids.\\
For negative delay times, no change in the harmonics is observed as the control pulse arrives after the generation process. When the control pulse starts to overlap with the generation pulse a strong suppression is observed. During pulse overlap the control pulse can directly affect the carrier dynamics during HHG. Significant suppression is also observable outside the direct overlap region for positive delay times, where for increasing delay time the suppression gradually recovers up to the unsuppressed HHG yield. For these delay times, the control pulse can not directly affect the HHG process but instead does this via the excitation of charge carriers. The presence of an initial carrier population generated by the control pulse is what causes the suppression. The carrier population will over time be affected by relaxation and recombination, which bring the excited charge carriers back to their ground state, this is directly observed in the recovery of harmonic yield for longer delay times. Increased suppression is found both for increasing control intensity and for increasing harmonic order. The fact that suppression increases with harmonic order strongly suggests that the underlying suppression mechanisms predominantly affect the interband current.\\

\section{Mechanisms of suppression of high-harmonic generation }
So far we have described the common observation of HHG suppression and its significant features, in the next we will discuss the specific microscopic mechanisms that can explain these observations. Four main possible mechanisms were identified, these being state blocking, excitation-induced dephasing (EID), insulator-to-metal phase transitions (IMT), and field modulation. Table \ref{table:overview} shows which of these mechanisms have been discussed in recent works where suppression was observed. In the following sections, we will discuss these mechanisms individually.

\begin{table}
  \caption{An overview of recent works showing suppression of HHG in various materials. The mechanisms indicate the suppression mechanisms discussed in these works, these being state blocking, excitation-induced dephasing (EID), insulator-to-metal phase transitions (IMT), and field modulation. $\lambda_{\text{gen.}}$ and $\lambda_{\text{con.}}$ respectively indicate the generation and control wavelength used. HO indicates the harmonic orders that have been shown from these experiments.\\
  *state blocking is explicitly discussed, however, is determined not to be the main cause of suppression.\\
  ** MAPbBr: Methylammonium lead bromide}
  \begin{tabular}{llccll}
    \hline
    Mechanism(s)  & Material & $\lambda_{\text{gen.}}$ [nm] & $\lambda_{\text{con.}}$ [nm] & HO &  \\
    \hline
    state blocking & ZnO & 3500 & 400  & 7, 11, 13 & Ref. \cite{Wang2017} \\
    state blocking & MoS$_2$ & 1560 to 2385 & 400  & 3, 4, 5  & Ref. \cite{Wang2022} \\
    state blocking & Graphene & 1350 & 400, 800  & 3  & Ref. \cite{Cheng2020} \\
    \hline
    EID & ZnO & 2350 & 400, 800  & 5, 7 & Ref. \cite{Xu2022}  \\
    EID & MoS$_2$ & 5000 & 660  & 5 to 16 & Ref. \cite{Heide2022} \\
    EID (*state blocking) & WSe$_2$ & 4770 & 760  & 5, 7 to 12& Ref. \cite{Nagai2023} \\
    EID & **MAPbBr & 1440 to 2320 & 400 & 3, 5, 7 & Ref. \cite{Geest23a} \\
    \hline
    IMT & VO$_2$ & 7000, 10000 & 1500  & 5, 7, 9  & Ref. \cite{Bionta2021} \\
    IMT & NbO$_2$ & 1800, 2160 & 400 & 3, 5  & Ref. \cite{nie23a} \\
    \hline
    field modulation & ZnO & 3500 & 1300  & 9, 11 & Ref. \cite{Wang2023} \\
    \hline
  \end{tabular}
  \label{table:overview}
\end{table}

\subsection{State blocking}
\begin{figure}[H]
    \centering
    \includegraphics[width=0.9\textwidth]{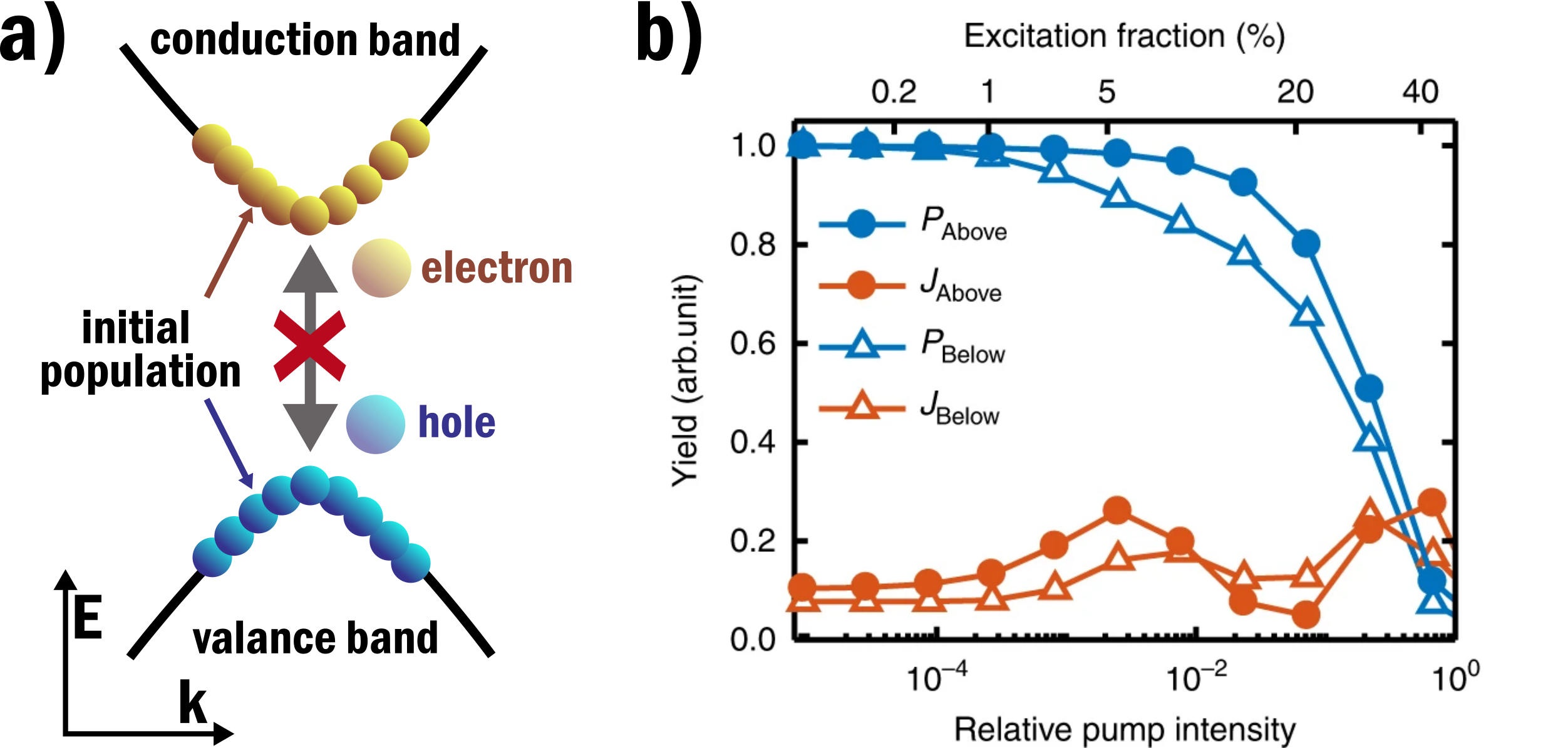}
    \caption{In (a) the effect of state blocking is schematically shown where the the generation of new electron-hole pairs is inhibited by the presence of an initial carrier population. In (b) SBE simulation results for the harmonic yield generated by a 3500 nm generation pulse with a 400 nm control pulse in ZnO, by Ref. \cite{Wang2017}. $P$ and $J$ respectively indicate the interband and intraband contributions and the yield is shown separately for the above and below bandgap harmonics. The yield is shown as a function of relative control intensity, as well as, the corresponding excitation fraction. The corresponding measurements in Ref. \cite{Wang2017} showed near complete suppression so for these simulations to match, excitation fractions over 40\% are required.  }
    \label{fig:stateBlocking}
\end{figure}

A number of works \cite{Wang2017,Wang2022,Cheng2020,Nagai2023} have looked into the effects of state blocking in order to explain the observed suppression. The idea here is that the control pulse generates an initial carrier population that occupies some of the excited states. The occupation of these excited states prevents the excitation of coherent electron-hole pairs to these states by the generation pulse. Similarly, the lack of electrons in the valance band will also inhibit excitation, this is referred to as ground state depletion. When the excitation by the generation pulse is prevented the HHG is suppressed. \\
The effects of state blocking and ground state depletion have been observed in a variety of different systems \cite{Margalit,Chong2010,Jannin2022,Kastner1992} which motivates the investigation of this mechanism in the context of HHG suppression.\\
The impact of state blocking has been investigated using simulations based on SBE. These SBE simulations make use of second-quantization to describe the electron dynamics which enables the simulation of state blocking. Different SBE simulations \cite{Wang2017,Nagai2023} show that the effects of state blocking are likely negligible for realistic parameters. In Fig. \ref{fig:stateBlocking} the results of an SBE simulations of HHG in ZnO are shown. 
In the corresponding experiments \cite{Wang2017}, near-complete suppression of the harmonics was observed which would require excitation fractions above 40\% to be consistent with these simulations. Realistic excitation fractions in semiconductors are considered to be only a few percent as above this dielectric breakdown occurs which results in material damage. 
The big discrepancy between simulations and experiments is a strong indication that state blocking by itself can not solely account for the significant observed suppression of HHG.\\

\subsection{Excitation-induced dephasing}
\begin{figure}[H]
    \centering
    \includegraphics[width=0.75\textwidth]{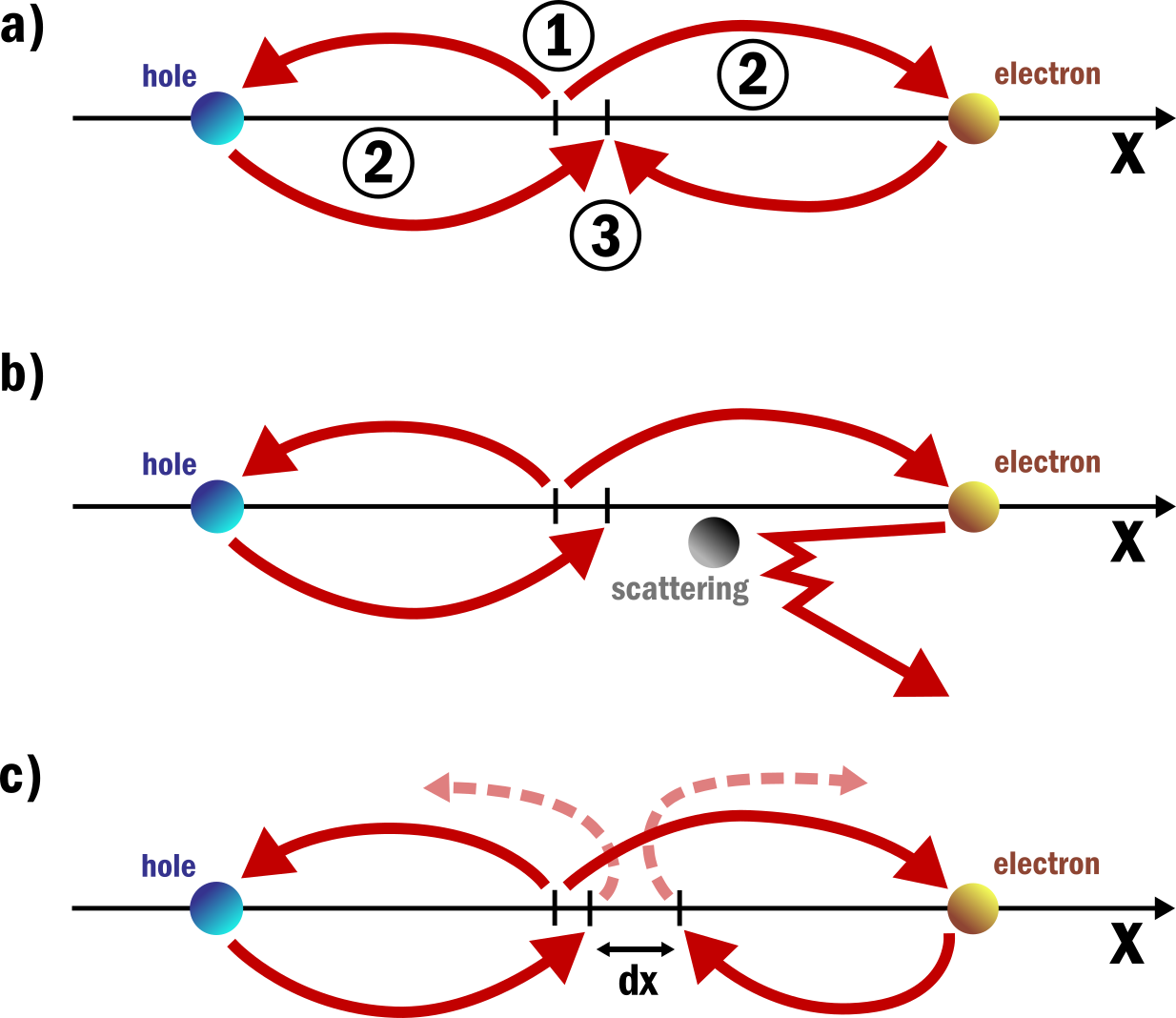}
    \caption{In (a) the the three-step model is schematically shown in real space. In step \textbf{1} an electron-hole pair is generated, in step \textbf{2} the charge carriers are accelerated, and in step \textbf{3} the electron and hole recombine. In (b) a scattering event of one of the charge carriers during the propagation is included. The scattering event causes the electron and hole to lose spatial coherence and prevents their coherent recombination. The scattering shown here can be carrier-carrier scattering, carrier-phonon scattering, or scattering of defects all of which will result in the loss of coherence. In (c) the electron and hole recombination is inhibited by an induced spatial displacement d$x$. This spatial displacement can be induced by having an additional electric field present during the acceleration of the electron-hole pair.}
    \label{fig:EIDandFieldModulation_RealSpace}
\end{figure}

Coherence between the generated electron-hole pair is essential for the effective recombination resulting in the interband current. To effectively illustrate this we can consider the semi-classical three-step model in real space as is shown in Fig. \ref{fig:EIDandFieldModulation_RealSpace}a. After excitation, the electron and hole are spatially separated as they are driven by the electric field. If during propagation one of the carriers scatters, as shown in Fig. \ref{fig:EIDandFieldModulation_RealSpace}b, part of its momentum and/or energy is transferred. As a result of this scattering the electron-hole pair will lose spatial coherence and will not coherently recombine. Figure \ref{fig:EIDandFieldModulation_RealSpace}c shows an alternative manner in which recombination can be inhibited, this figure will be discussed in more detail in the section on field modulation. The loss of spatial coherence is also referred to as dephasing. Dephasing can be caused by all different scattering events including carrier-carrier scattering, carrier-phonon scattering, and scattering of defects.\\ 
The importance of dephasing has consistently been confirmed by the modeling of HHG. In models, the dephasing effectively functions as a damping term that stops the build-up of the coherent population over multiple optical cycles. As the particle density of solids greatly exceeds that of gases dephasing plays a significantly greater role.\\
In order to match simulations of HHG from solids with experiments, short dephasing times of only a few femtoseconds have to be used \cite{Vampa2014,Yue2021}. This is considerably lower than the dephasing time found with photon-echo measurements, where dephasing times are closer to tens of femtoseconds \cite{Becker1988}. By including the macroscopic propagation it is possible to explain the discrepancy between dephasing times \cite{Kilen2020}. This propagation-induced dephasing effect does however require sample thicknesses of at least a few micrometers and thus can not explain the short dephasing times of thin and monolayer samples. More recently, by considering dephasing in real space the short dephasing times were attributed to recombination events of carriers that experience large spatial separation \cite{Brown2022}. We also note that for HHG the carriers are accelerated to much higher momenta than in photon-echo measurements. Therefore we see the significant discrepancy in dephasing times as a strong indication that the dephasing time is strongly dependent on the carrier momentum. \\
The excitation of carriers by the control pulse can cause an alteration of the dephasing time by increasing the likelihood of scattering. This is either directly by the presence of more excited carriers or mediated via phonon coupling. A reasonably lowered dephasing time can account for the significant suppression observed in experiments \cite{Heide2022,Nagai2023}. Moreover, the observation of strong harmonic dependence when it comes to suppression is a convincing sign that dephasing dominates high-harmonic suppression. This can intuitively be understood as higher harmonics require longer acceleration times and thus will be affected more when the dephasing rate is increased \cite{Heide2022}.\\
Control over the dephasing rate in solids will thus enable major control over the high-harmonic generation in a material. Interestingly, a reduction of the dephasing rate should allow for a significant increase in high-harmonic yield.\\
The difficulty with dephasing is that it results from carrier scattering, which is intrinsically a multi-electron effect, as well as carrier-phonon scattering, which requires an accurate description of electron-phonon coupling. Photon-echo experiments \cite{Becker1988} have measured photocarrier-density dependent dephasing times that are linked to carrier scattering, but only at low excitation intensities, for single-photon absorption at the $\Gamma$ point - all conditions that are not fulfilled in HHG.  
Current state-of-the-art simulations either make use of SBE or density functional theory (DFT). In SBE simulation dephasing is added via a phenomenological damping term to the EOM, as is also shown in equation \ref{eq:EOM}\cite{Yue2021}. $T_2$ is the dephasing time while $T_1$ refers to the recombination time. Conventionally in SBE simulations, $T_2$ is chosen to accomplish good agreement between simulation results and measurements \cite{Yue2021,vampa14a}. This means that SBE simulations that do not explicitly calculate $T_2$ have no predictive power when it comes to the dephasing time. 
Time-dependent DFT simulations have been applied to solid HHG \cite{tancogne17a} and allow for the inclusion of some scattering effects depending on the choice of the exchange-correlation functional. The complexity of DFT simulations, however, can make gaining physical intuition challenging, and the multielectron nature of $T_2$ makes it particularly difficult to grasp in DFT. Moreover, some dephasing effects are not easily implemented in DFT simulations such as the contribution of material defects.\\
New advancements in modeling and understanding the scattering in solids as causing dephasing in high-harmonic generation will be key to the advancement of all-optical high-harmonic modulation.

\subsection{Insulator-to-metal phase transitions}
\begin{figure}[H]
    \centering
    \includegraphics[width=0.6\textwidth]{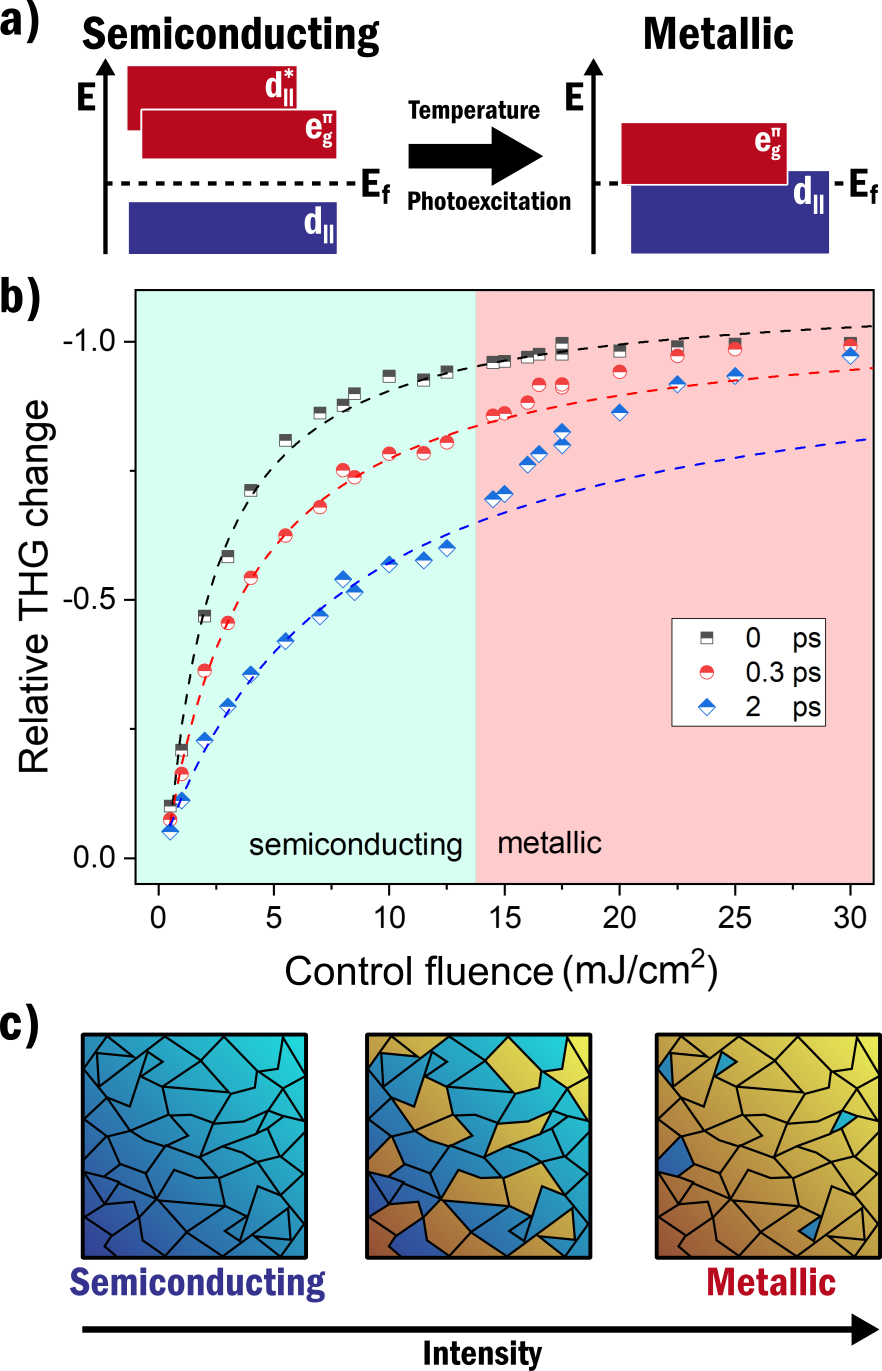}
    \caption{(a) Restructuring of the energy states around the bandgap of NbO$_2$ when it undergoes the insulator-to-metal phase transition. This phase transition can be initialized by temperature (around $1080$ K) or by photoexcitation. In the semiconducting state, NbO$_2$ has a bandgap of around 0.7-1.1 eV. In (b) the suppression of the third harmonic from NbO$_2$ is shown for increasing control intensity for three different pulse delays measured by Ref. \cite{nie23a}. 1800 nm generation pulses were used in combination with 400 nm control pulses, and the measurements were performed at room temperature. A clear deviation from the exponential suppression is observed at around 13.5 mJ/cm$^2$ indicating the material phase transition. In (c) the spatial states of NbO$_2$ are shown for increased illumination intensity. Segments of semiconducting and metallic states can exist within a NbO$_2$ sample at the same time. To study individual segments high-resolution spatially resolved measurements are required.  } 
    \label{fig:CEM}
\end{figure}

Both state blocking and excitation-induced dephasing focus on carrier interactions and do not consider significant electronic potential changes in the material. For most conventional materials carrier excitation has a limited effect on the electronic structure, with effects such as bandgap renormalization being small \cite{bennett90a}.\\
Contrary to this are strongly correlated materials (SCMs) which can undergo phase transition under carrier excitation \cite{imada1998metal, wegkamp2015ultrafast, wegkamp2014instantaneous, wang2022coherent}. Significant changes to the electronic structure will have a major impact on the carrier dynamic and as a consequence also on the HHG emission. In this section, we will focus on a particular class of SCMs, where the insulator-to-metal phase transition (IMT) has been found, and also studied via HHG.\cite{Bionta2021,nie23a}\\
IMT in SCMs refers to the material switch from the insulating or semiconducting phase to the metallic phase under certain external triggers, such as temperature or photo-excitation \cite{wegkamp2015ultrafast, wegkamp2014instantaneous, wang2022coherent}. As an example, Fig. \ref{fig:CEM}a schematically shows the IMT in NbO$_2$ where the merging of the separated orbitals around the fermi level results in the collapse of the band gap \cite{wegkamp2015ultrafast}. 
Due to their unique electronic properties, SCMs have been of significant interest to the development of new novel devices. For example, IMT materials are a key enabler of Mott memristors which in turn pave the way for neuromorphic computing \cite{yang2011oxide, kumar2020third}. While IMT in SCMs has been studied over the last decade, a number of relevant open questions remain. For example, how to distinguish between the competing electron-electron and electron-lattice interactions in the IMT.\\
Due to its highly nonlinear nature, HHG is extremely sensitive to the IMT, as HHG emission can be greatly altered by any subtle change in electronic or lattice structures, let alone the bandgap collapse. 
It is possible to directly observe the IMT via time-resolved HHG. Figure \ref{fig:CEM}b shows the IMT in NbO$_2$ at a control fluence of 12 mJ/cm$^2$ via the clear deviation from a saturation model \cite{nie23a}. The saturation model considers the suppression due to lowered dephasing time in the semiconducting state.
The deviation corresponds to a sudden increased suppression of HHG and is assigned to photo-induced IMT. \\
To explain the greatly reduced HHG efficiency in the metallic phase two main effects can be considered: firstly, a higher density of free carriers in the metallic phase results in a significantly reduced dephasing time; secondly, the bandgap collapse results in carriers movement closer to that of free carriers \cite{Korobenko2021}. Free carriers miss the non-linear response to the driving field required for generating high-harmonics. Both effects together can result in a very strong, practically complete signal suppression, as shown in Fig. \ref{fig:CEM}b.\\
Besides all-optical control of HHG, this very strong signal suppression enables HHG as an ultra-sensitive probe for ultra-fast nanoscopy in SCMs. As the HHG process takes place within an optical cycle it becomes possible to resolve the phase transitions with very high temporal resolution of only a few femtoseconds and possible sub-femtosecond. This will enable the identification of electronic contributions in IMT.\\ 
Except for the temporal properties of IMT, spatial information could also be accessed using HHG imaging. As shown in Fig. \ref{fig:CEM}c the IMT will not occur throughout the whole material at the same time. Spatial imaging of the IMT will benefit greatly from the aforementioned super-resolution imaging techniques which make use of the HHG modulation and have already been demonstrated to work for NbO$_2$ \cite{Murzyn2023}.\\
Optical modulation of HHG in solids is promising in simultaneously providing the temporal and spatial information of IMT, which not only benefits a comprehensive understanding of IMTs but can also guide the device design with SCMs. 

\subsection{Field modulation} 
\begin{figure}[H]
    \centering
    \includegraphics[width=0.9\textwidth]{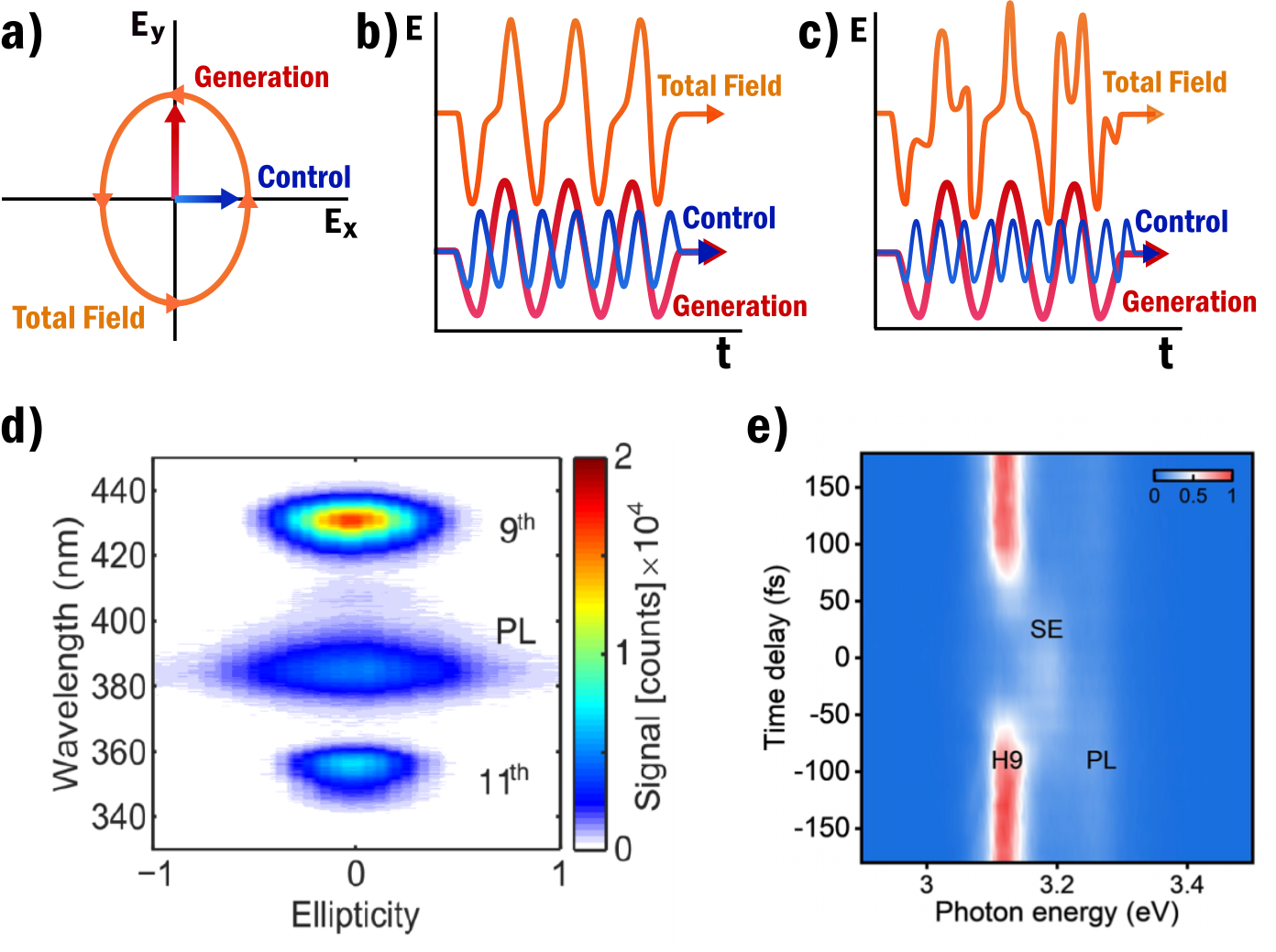}
    \caption{
    In (a), (b), and (c) the different frequency relations between the generation $\omega_g$ and control $\omega_c$ pulse are shown. In (a) the generation and control frequencies are degenerate which means that substantial field changes can only be accomplished when the fields are out-of-phase and orthogonal such that elliptical polarization can be achieved. In (b) the control frequency is an integer multiple of the generation frequency which results in equal modulation of every optical cycle. In (c) the control frequency is not an integer multiple of the generation frequency such that every optical cycle is modulated differently. In (b) and (c) the total field curve has been offset to better show its shape. In (d) ZnO HHG results from Ref. \cite{Hollinger2021} are shown for different generation ellipticity. The ellipticity of -1 and 1 corresponds to the left- and right-handed circular polarization while 0 corresponds to linear polarization. PL indicates the photoluminescence. In (e) ZnO HHG results from Ref. \cite{Wang2023} are shown for 3500 nm generation and 1300 nm control wavelengths. As the control wavelength far exceeds the bandgap of ZnO (around 368 nm \cite{Rodnyi2011,Xu2022}) suppression can only be observed in the region of pulse overlap. PL indicates the photoluminescence and SE indicates some stimulated emission signal.  }
    \label{fig:Modulation}
\end{figure}

So far we have discussed mechanisms where HHG was affected via carrier excitation of the control pulse. When there is overlap between the generation and control pulse the field of the control pulse can affect the electron dynamics of the HHG process directly. During overlap, we can consider the control pulse as a modulation of the generation field.\\
Shaped multi-colour fields have been demonstrated to allow for an increased HHG yield, an increased cut-off frequency, and divergence control in gases \cite{Li2019,Kroh2018,RoscamAbbing2020,roscam21a}. In these cases, the field is shaped such that more of the excited charge carriers are able to coherently recombine by tuning the excitation rate and trajectories. While these works focus on increasing the HHG yield the same principle can be used to achieve suppression by instead inhibiting coherent recombination. We can imagine this suppression in real space by considering that the control field induces a spatial displacement between carriers, as shown in Fig. \ref{fig:EIDandFieldModulation_RealSpace}c. The control field can affect the trajectories such that the induced spatial displacement inhibits recombination and as a result, suppresses the harmonics. Alternatively viewed, the control field can modulate the carrier movement through reciprocal space which allows it to impact the intraband current. Additionally, for strong enough control fields, the time at which significant excitation occurs can also be altered. \\
Depending on the wavelength ratio between the generation and control pulse various modulations of the electric field are possible. One way of modulating the generation pulse is by having it become elliptically polarised, this is equivalent to adding a degenerate out-of-phase orthogonal component to the generation field, as is also shown in Fig. \ref{fig:Modulation}a. In practice, elliptical polarization is achieved by making use of quarter-waveplates. For various materials, near-complete HHG suppression is observed when ellipticity is increased \cite{Liu2016,Hollinger2021,Yoshikawa2017}, an example of which is shown in Fig. \ref{fig:Modulation}d where also a clear harmonic dependence is observed.\\
To effectively control the field shape during an optical cycle non-degenerate wavelengths have to be used. To modulate all optical cycles equally the control frequency has to be an integer multiple of the generation frequency, as shown in Fig. \ref{fig:Modulation}b, if this is not the case all optical cycles are modulated differently as shown in Fig. \ref{fig:Modulation}c. For the aforementioned multi-color HHG process second and third harmonic generations are used to allow for consistent modulation of the optical cycles. For efficient suppression, this consistent modulation is not necessary as is shown by the results in Fig. \ref{fig:Modulation}e where near complete suppression is observed by using a non-commensurate control pulse. Note that, the suppression is only observed in the region of overlap as the control pulse in this experiment is far enough below the bandgap such that it does not excite carriers. \\
Modulation and suppression have mostly been found to be the most significant during overlap, as a result, applications utilizing HHG suppression will likely rely on processes governed by the effect of field modulation. As there is no requirement for exciting charge carriers there is a lessened importance of the material and an increased importance of the electric field. The loosening of this requirement also allows the use of higher wavelength and lower intensity control pulses for achieving significant suppression.\\
As the modulation in the system is caused just by the electric field, simulations of these systems are much more feasible compared to those considering dephasing where scattering interactions have to be considered. An exciting challenge lies in the optimization of the control pulse for maximum modulation of high-harmonic generation, certainly considering the great number of possible parameters to optimize.

\section{Conclusion and outlook}
In this paper, the microscopic mechanisms underlying the suppression of HHG in solids have been discussed. Four main mechanisms were identified, these being: state blocking, excitation-induced dephasing, insulator-to-metal phase transitions, and field modulation.\\
Simulations of state blocking indicated only a minor contribution to the suppression of HHG in conventional semiconductors, which indicates that state blocking by itself can not account for the significant suppression observed in recent measurements.\\
Excitation-induced dephasing can account for the significant suppression observed as well as the further temporal dynamics observed in experiments. The scattering underlying dephasing makes it difficult to model and exceeds the limits of current state-of-the-art simulations. An interesting challenge lies in enabling predictive modeling of the dephasing, which would, in turn, enable the precise study of the underlying electron and phonon dynamics.\\
For IMT materials a significantly increased HHG suppression occurs due to the bandgap collapse. Due to the possibility for high temporal and spatial resolution, HHG provides an exciting platform for resolving the material phase transitions, which will greatly benefit the development of SCMs devices.\\
The most significant suppression is consistently found during pulse overlap where the control field effectively modulates the generation field. The significance of the observed suppression makes this mechanism especially relevant for applications such as all-optical signal modulation and super-resolution imaging. The significant degrees of freedom with regard to generation and control pulse combinations open the door for precise field optimization to obtain a very controlled modulation of HHG.\\
Improving the understanding of these microscopic mechanisms underpinning the HHG process will be key to achieving complete optical control of the HHG process. Such complete all-optical control has numerous applications.  We have elaborated on potential for label-free super resolution microscopy via deactivated high-harmonic generation in solids. 
Additionally, by controlling the carrier excitation during the HHG process, sub-diffraction controlled chemistry on the nanoscale may become possible. This seems particularly feasible for field modulation, where experience from gas HHG has shown that the excitation step of HHG can be controlled effectively \cite{RoscamAbbing2020,roscam21a}. Finally, nonlinear solid light sources are explored in many fields, for example via nonlinear dielectric metasurfaces \cite{2018Kivshar} that offer great deals of beam control via nanoscale engineering of structures. Optical control of HHG adds another control knob that enables femtosecond switching of these metasurfaces. Consequently, we are convinced that optical control of HHG has numerous applications in the future, as it represents a fully controllable femtosecond, possibly even sub-femtosecond, all-optical switch.

\begin{acknowledgement}
This work has been carried out at the Advanced Research Center for Nanolithography (ARCNL), a public-private partnership of the University of Amsterdam (UvA), the Vrije Universiteit Amsterdam (VU), the Netherlands Organisation for Scientific Research (NWO), and the semiconductor equipment manufacturer ASML, and was partly financed by ‘Toeslag voor Topconsortia voor Kennis en Innovatie (TKI)’ from the Dutch Ministry of Economic Affairs and Climate Policy. 
This manuscript is part of a project that has received funding from the European Research Council (ERC) under the European Union’s Horizon Europe research and innovation programme (grant agreement no. 101041819, ERC Starting Grant ANACONDA) and funded P.v.E and partly P.K.
The manuscript is also part of the VIDI research programme HIMALAYA with project number VI.Vidi.223.133 financed by NWO, which partly funded P.K.
Z.N., and P.M.K. acknowledge support from the Open Technology Programme (OTP) by NWO, grant no. 18703.
\end{acknowledgement}
\bibliography{achemso-demo}

\providecommand{\latin}[1]{#1}
\makeatletter
\providecommand{\doi}
  {\begingroup\let\do\@makeother\dospecials
  \catcode`\{=1 \catcode`\}=2 \doi@aux}
\providecommand{\doi@aux}[1]{\endgroup\texttt{#1}}
\makeatother
\providecommand*\mcitethebibliography{\thebibliography}
\csname @ifundefined\endcsname{endmcitethebibliography}
  {\let\endmcitethebibliography\endthebibliography}{}
\begin{mcitethebibliography}{85}
\providecommand*\natexlab[1]{#1}
\providecommand*\mciteSetBstSublistMode[1]{}
\providecommand*\mciteSetBstMaxWidthForm[2]{}
\providecommand*\mciteBstWouldAddEndPuncttrue
  {\def\EndOfBibitem{\unskip.}}
\providecommand*\mciteBstWouldAddEndPunctfalse
  {\let\EndOfBibitem\relax}
\providecommand*\mciteSetBstMidEndSepPunct[3]{}
\providecommand*\mciteSetBstSublistLabelBeginEnd[3]{}
\providecommand*\EndOfBibitem{}
\mciteSetBstSublistMode{f}
\mciteSetBstMaxWidthForm{subitem}{(\alph{mcitesubitemcount})}
\mciteSetBstSublistLabelBeginEnd
  {\mcitemaxwidthsubitemform\space}
  {\relax}
  {\relax}

\bibitem[{Corkum} and {Krausz}(2007){Corkum}, and {Krausz}]{2007Corkum}
{Corkum},~P.~B.; {Krausz},~F. {Attosecond science}. \emph{Nature Physics}
  \textbf{2007}, \emph{3}, 381--387\relax
\mciteBstWouldAddEndPuncttrue
\mciteSetBstMidEndSepPunct{\mcitedefaultmidpunct}
{\mcitedefaultendpunct}{\mcitedefaultseppunct}\relax
\EndOfBibitem
\bibitem[Calegari \latin{et~al.}(2016)Calegari, Sansone, Stagira, Vozzi, and
  Nisoli]{Calegari2016}
Calegari,~F.; Sansone,~G.; Stagira,~S.; Vozzi,~C.; Nisoli,~M. Advances in
  attosecond science. \emph{Journal of Physics B: Atomic, Molecular and Optical
  Physics} \textbf{2016}, \emph{49}, 062001\relax
\mciteBstWouldAddEndPuncttrue
\mciteSetBstMidEndSepPunct{\mcitedefaultmidpunct}
{\mcitedefaultendpunct}{\mcitedefaultseppunct}\relax
\EndOfBibitem
\bibitem[Li \latin{et~al.}(2020)Li, Lu, Chew, Han, Li, Wu, Wang, Ghimire, and
  Chang]{Li2020}
Li,~J.; Lu,~J.; Chew,~A.; Han,~S.; Li,~J.; Wu,~Y.; Wang,~H.; Ghimire,~S.;
  Chang,~Z. Attosecond science based on high harmonic generation from gases and
  solids. \emph{Nature Communications} \textbf{2020}, \emph{11}, 2748\relax
\mciteBstWouldAddEndPuncttrue
\mciteSetBstMidEndSepPunct{\mcitedefaultmidpunct}
{\mcitedefaultendpunct}{\mcitedefaultseppunct}\relax
\EndOfBibitem
\bibitem[Vampa and Brabec(2017)Vampa, and Brabec]{Vampa2017}
Vampa,~G.; Brabec,~T. Merge of high harmonic generation from gases and solids
  and its implications for attosecond science. \emph{Journal of Physics B:
  Atomic, Molecular and Optical Physics} \textbf{2017}, \emph{50}, 083001\relax
\mciteBstWouldAddEndPuncttrue
\mciteSetBstMidEndSepPunct{\mcitedefaultmidpunct}
{\mcitedefaultendpunct}{\mcitedefaultseppunct}\relax
\EndOfBibitem
\bibitem[Kraus \latin{et~al.}(2018)Kraus, Z{\"u}rch, Cushing, Neumark, and
  Leone]{kraus18a}
Kraus,~P.~M.; Z{\"u}rch,~M.; Cushing,~S.~K.; Neumark,~D.~M.; Leone,~S.~R. The
  ultrafast X-ray spectroscopic revolution in chemical dynamics. \emph{Nature
  Reviews Chemistry} \textbf{2018}, \emph{2}, 82\relax
\mciteBstWouldAddEndPuncttrue
\mciteSetBstMidEndSepPunct{\mcitedefaultmidpunct}
{\mcitedefaultendpunct}{\mcitedefaultseppunct}\relax
\EndOfBibitem
\bibitem[Kraus and W{\"o}rner(2018)Kraus, and W{\"o}rner]{kraus18b}
Kraus,~P.~M.; W{\"o}rner,~H.~J. Perspectives of attosecond spectroscopy for the
  understanding of fundamental electron correlations. \emph{Angewandte Chemie
  International Edition} \textbf{2018}, \emph{57}, 5228--5247\relax
\mciteBstWouldAddEndPuncttrue
\mciteSetBstMidEndSepPunct{\mcitedefaultmidpunct}
{\mcitedefaultendpunct}{\mcitedefaultseppunct}\relax
\EndOfBibitem
\bibitem[W{\"o}rner \latin{et~al.}(2017)W{\"o}rner, Arrell, Banerji, Cannizzo,
  Chergui, Das, Hamm, Keller, Kraus, Liberatore, \latin{et~al.}
  others]{woerner17a}
W{\"o}rner,~H.~J.; Arrell,~C.~A.; Banerji,~N.; Cannizzo,~A.; Chergui,~M.;
  Das,~A.~K.; Hamm,~P.; Keller,~U.; Kraus,~P.~M.; Liberatore,~E.,
  \latin{et~al.}  Charge migration and charge transfer in molecular systems.
  \emph{Structural dynamics} \textbf{2017}, \emph{4}, 061508\relax
\mciteBstWouldAddEndPuncttrue
\mciteSetBstMidEndSepPunct{\mcitedefaultmidpunct}
{\mcitedefaultendpunct}{\mcitedefaultseppunct}\relax
\EndOfBibitem
\bibitem[Jansen \latin{et~al.}(2018)Jansen, de~Beurs, Liu, Eikema, and
  Witte]{jansen18a}
Jansen,~G. S.~M.; de~Beurs,~A.; Liu,~X.; Eikema,~K. S.~E.; Witte,~S.
  Diffractive shear interferometry for extreme ultraviolet high-resolution
  lensless imaging. \emph{Opt. Express} \textbf{2018}, \emph{26},
  12479--12489\relax
\mciteBstWouldAddEndPuncttrue
\mciteSetBstMidEndSepPunct{\mcitedefaultmidpunct}
{\mcitedefaultendpunct}{\mcitedefaultseppunct}\relax
\EndOfBibitem
\bibitem[Loetgering \latin{et~al.}(2021)Loetgering, Liu, De~Beurs, Du, Kuijper,
  Eikema, and Witte]{loetgering2021tailoring}
Loetgering,~L.; Liu,~X.; De~Beurs,~A.~C.; Du,~M.; Kuijper,~G.; Eikema,~K.~S.;
  Witte,~S. Tailoring spatial entropy in extreme ultraviolet focused beams for
  multispectral ptychography. \emph{Optica} \textbf{2021}, \emph{8},
  130--138\relax
\mciteBstWouldAddEndPuncttrue
\mciteSetBstMidEndSepPunct{\mcitedefaultmidpunct}
{\mcitedefaultendpunct}{\mcitedefaultseppunct}\relax
\EndOfBibitem
\bibitem[Porter \latin{et~al.}(2023)Porter, Coenen, Geypen, Scholz, van
  Rijswijk, Nienhuys, Ploegmakers, Reinink, Cramer, van Laarhoven,
  \latin{et~al.} others]{porter23a}
Porter,~C.; Coenen,~T.; Geypen,~N.; Scholz,~S.; van Rijswijk,~L.;
  Nienhuys,~H.-K.; Ploegmakers,~J.; Reinink,~J.; Cramer,~H.; van Laarhoven,~R.,
  \latin{et~al.}  Soft x-ray: novel metrology for 3D profilometry and device
  pitch overlay. Metrology, Inspection, and Process Control XXXVII. 2023; pp
  412--420\relax
\mciteBstWouldAddEndPuncttrue
\mciteSetBstMidEndSepPunct{\mcitedefaultmidpunct}
{\mcitedefaultendpunct}{\mcitedefaultseppunct}\relax
\EndOfBibitem
\bibitem[Goulielmakis and Brabec(2022)Goulielmakis, and
  Brabec]{Goulielmakis2022}
Goulielmakis,~E.; Brabec,~T. High harmonic generation in condensed matter.
  \emph{Nature Photonics} \textbf{2022}, \emph{16}, 411--421\relax
\mciteBstWouldAddEndPuncttrue
\mciteSetBstMidEndSepPunct{\mcitedefaultmidpunct}
{\mcitedefaultendpunct}{\mcitedefaultseppunct}\relax
\EndOfBibitem
\bibitem[Huttner \latin{et~al.}(2016)Huttner, Schuh, Moloney, and
  Koch]{Huttner2016}
Huttner,~U.; Schuh,~K.; Moloney,~J.~V.; Koch,~S.~W. Similarities and
  differences between high-harmonic generation in atoms and solids.
  \emph{Journal of the Optical Society of America B} \textbf{2016}, \emph{33},
  C22\relax
\mciteBstWouldAddEndPuncttrue
\mciteSetBstMidEndSepPunct{\mcitedefaultmidpunct}
{\mcitedefaultendpunct}{\mcitedefaultseppunct}\relax
\EndOfBibitem
\bibitem[Ghimire \latin{et~al.}(2011)Ghimire, Dichiara, Sistrunk, Agostini,
  Dimauro, and Reis]{Ghimire2011}
Ghimire,~S.; Dichiara,~A.~D.; Sistrunk,~E.; Agostini,~P.; Dimauro,~L.~F.;
  Reis,~D.~A. Observation of high-order harmonic generation in a bulk crystal.
  \emph{Nature Physics} \textbf{2011}, \emph{7}, 138--141\relax
\mciteBstWouldAddEndPuncttrue
\mciteSetBstMidEndSepPunct{\mcitedefaultmidpunct}
{\mcitedefaultendpunct}{\mcitedefaultseppunct}\relax
\EndOfBibitem
\bibitem[Vampa \latin{et~al.}(2016)Vampa, Hammond, Thiré, Schmidt, Légaré,
  Klug, and Corkum]{Vampa2016}
Vampa,~G.; Hammond,~T.~J.; Thiré,~N.; Schmidt,~B.~E.; Légaré,~F.;
  Klug,~D.~D.; Corkum,~P.~B. Generation of high harmonics from silicon.
  \emph{arXiv:1605.06345} \textbf{2016}, \relax
\mciteBstWouldAddEndPunctfalse
\mciteSetBstMidEndSepPunct{\mcitedefaultmidpunct}
{}{\mcitedefaultseppunct}\relax
\EndOfBibitem
\bibitem[Luu \latin{et~al.}(2015)Luu, Garg, Kruchinin, Moulet, Hassan, and
  Goulielmakis]{Luu2015}
Luu,~T.~T.; Garg,~M.; Kruchinin,~S.~Y.; Moulet,~A.; Hassan,~M.~T.;
  Goulielmakis,~E. Extreme ultraviolet high-harmonic spectroscopy of solids.
  \emph{Nature} \textbf{2015}, \emph{521}, 498--502\relax
\mciteBstWouldAddEndPuncttrue
\mciteSetBstMidEndSepPunct{\mcitedefaultmidpunct}
{\mcitedefaultendpunct}{\mcitedefaultseppunct}\relax
\EndOfBibitem
\bibitem[You \latin{et~al.}(2017)You, Reis, and Ghimire]{You2017}
You,~Y.~S.; Reis,~D.~A.; Ghimire,~S. Anisotropic high-harmonic generation in
  bulk crystals. \emph{Nature Physics} \textbf{2017}, \emph{13}, 345--349\relax
\mciteBstWouldAddEndPuncttrue
\mciteSetBstMidEndSepPunct{\mcitedefaultmidpunct}
{\mcitedefaultendpunct}{\mcitedefaultseppunct}\relax
\EndOfBibitem
\bibitem[Wang \latin{et~al.}(2017)Wang, Park, Lai, Xu, Blaga, Yang, Agostini,
  and DiMauro]{Wang2017}
Wang,~Z.; Park,~H.; Lai,~Y.~H.; Xu,~J.; Blaga,~C.~I.; Yang,~F.; Agostini,~P.;
  DiMauro,~L.~F. The roles of photo-carrier doping and driving wavelength in
  high harmonic generation from a semiconductor. \emph{Nature Communications}
  \textbf{2017}, \emph{8}, 1686\relax
\mciteBstWouldAddEndPuncttrue
\mciteSetBstMidEndSepPunct{\mcitedefaultmidpunct}
{\mcitedefaultendpunct}{\mcitedefaultseppunct}\relax
\EndOfBibitem
\bibitem[Xu \latin{et~al.}(2022)Xu, Zhang, Yu, Han, Wang, and Hu]{Xu2022}
Xu,~S.; Zhang,~H.; Yu,~J.; Han,~Y.; Wang,~Z.; Hu,~J. Ultrafast modulation of a
  high harmonic generation in a bulk ZnO single crystal. \emph{Optics Express}
  \textbf{2022}, \emph{30}, 41350\relax
\mciteBstWouldAddEndPuncttrue
\mciteSetBstMidEndSepPunct{\mcitedefaultmidpunct}
{\mcitedefaultendpunct}{\mcitedefaultseppunct}\relax
\EndOfBibitem
\bibitem[Wang \latin{et~al.}(2023)Wang, Liu, Jiang, Gao, Yang, Peng, Gong, and
  Wu]{Wang2023}
Wang,~Y.; Liu,~Y.; Jiang,~P.; Gao,~Y.; Yang,~H.; Peng,~L.-Y.; Gong,~Q.; Wu,~C.
  Optical switch of electron-hole and electron-electron collisions in
  semiconductors. \emph{Physical Review B} \textbf{2023}, \emph{107},
  L161301\relax
\mciteBstWouldAddEndPuncttrue
\mciteSetBstMidEndSepPunct{\mcitedefaultmidpunct}
{\mcitedefaultendpunct}{\mcitedefaultseppunct}\relax
\EndOfBibitem
\bibitem[Cheng \latin{et~al.}(2020)Cheng, Hong, Zhao, Wu, Pan, Liu, Zuo, Zhang,
  Xie, Wang, Yu, Ye, Meng, and Liu]{Cheng2020}
Cheng,~Y.; Hong,~H.; Zhao,~H.; Wu,~C.; Pan,~Y.; Liu,~C.; Zuo,~Y.; Zhang,~Z.;
  Xie,~J.; Wang,~J.; Yu,~D.; Ye,~Y.; Meng,~S.; Liu,~K. Ultrafast optical
  modulation of harmonic generation in two-dimensional materials. \emph{Nano
  Letters} \textbf{2020}, \emph{20}, 8053--8058\relax
\mciteBstWouldAddEndPuncttrue
\mciteSetBstMidEndSepPunct{\mcitedefaultmidpunct}
{\mcitedefaultendpunct}{\mcitedefaultseppunct}\relax
\EndOfBibitem
\bibitem[Yoshikawa \latin{et~al.}(2017)Yoshikawa, Tamaya, and
  Tanaka]{Yoshikawa2017}
Yoshikawa,~N.; Tamaya,~T.; Tanaka,~K. Optics: High-harmonic generation in
  graphene enhanced by elliptically polarized light excitation. \emph{Science}
  \textbf{2017}, \emph{356}, 736--738\relax
\mciteBstWouldAddEndPuncttrue
\mciteSetBstMidEndSepPunct{\mcitedefaultmidpunct}
{\mcitedefaultendpunct}{\mcitedefaultseppunct}\relax
\EndOfBibitem
\bibitem[Wang \latin{et~al.}(2022)Wang, Iyikanat, Bai, Hu, Das, Dai, Zhang, Du,
  Li, Lipsanen, Abajo, and Sun]{Wang2022}
Wang,~Y.; Iyikanat,~F.; Bai,~X.; Hu,~X.; Das,~S.; Dai,~Y.; Zhang,~Y.; Du,~L.;
  Li,~S.; Lipsanen,~H.; Abajo,~F. J. G.~D.; Sun,~Z. Optical Control of
  High-Harmonic Generation at the Atomic Thickness. \emph{Nano Letters}
  \textbf{2022}, \emph{22}, 8455--8462\relax
\mciteBstWouldAddEndPuncttrue
\mciteSetBstMidEndSepPunct{\mcitedefaultmidpunct}
{\mcitedefaultendpunct}{\mcitedefaultseppunct}\relax
\EndOfBibitem
\bibitem[Heide \latin{et~al.}(2022)Heide, Kobayashi, Johnson, Liu, Heinz, Reis,
  and Ghimire]{Heide2022}
Heide,~C.; Kobayashi,~Y.; Johnson,~A.~C.; Liu,~F.; Heinz,~T.~F.; Reis,~D.~A.;
  Ghimire,~S. Probing electron-hole coherence in strongly driven 2D materials
  using high-harmonic generation. \emph{Optica} \textbf{2022}, \emph{9},
  512\relax
\mciteBstWouldAddEndPuncttrue
\mciteSetBstMidEndSepPunct{\mcitedefaultmidpunct}
{\mcitedefaultendpunct}{\mcitedefaultseppunct}\relax
\EndOfBibitem
\bibitem[Nagai \latin{et~al.}(2023)Nagai, Uchida, Kusaba, Endo, Miyata, and
  Tanaka]{Nagai2023}
Nagai,~K.; Uchida,~K.; Kusaba,~S.; Endo,~T.; Miyata,~Y.; Tanaka,~K. Effect of
  incoherent electron-hole pairs on high harmonic generation in an atomically
  thin semiconductor. \emph{Physical Review Research} \textbf{2023}, \emph{5},
  043130\relax
\mciteBstWouldAddEndPuncttrue
\mciteSetBstMidEndSepPunct{\mcitedefaultmidpunct}
{\mcitedefaultendpunct}{\mcitedefaultseppunct}\relax
\EndOfBibitem
\bibitem[Bionta \latin{et~al.}(2021)Bionta, Haddad, Leblanc, Gruson, Lassonde,
  Ibrahim, Chaillou, Émond, Otto, Álvaro Jiménez-Galán, Silva, Ivanov,
  Siwick, Chaker, and Légaré]{Bionta2021}
Bionta,~M.~R.; Haddad,~E.; Leblanc,~A.; Gruson,~V.; Lassonde,~P.; Ibrahim,~H.;
  Chaillou,~J.; Émond,~N.; Otto,~M.~R.; Álvaro Jiménez-Galán,;
  Silva,~R.~E.; Ivanov,~M.; Siwick,~B.~J.; Chaker,~M.; Légaré,~F. Tracking
  ultrafast solid-state dynamics using high harmonic spectroscopy.
  \emph{Physical Review Research} \textbf{2021}, \emph{3}, 023250\relax
\mciteBstWouldAddEndPuncttrue
\mciteSetBstMidEndSepPunct{\mcitedefaultmidpunct}
{\mcitedefaultendpunct}{\mcitedefaultseppunct}\relax
\EndOfBibitem
\bibitem[Nie \latin{et~al.}(2023)Nie, Guery, Molinero, Juergens, van~den
  Hooven, Wang, Jimenez~Galan, Planken, Silva, and Kraus]{nie23a}
Nie,~Z.; Guery,~L.; Molinero,~E.~B.; Juergens,~P.; van~den Hooven,~T.~J.;
  Wang,~Y.; Jimenez~Galan,~A.; Planken,~P. C.~M.; Silva,~R. E.~F.; Kraus,~P.~M.
  Following the Nonthermal Phase Transition in Niobium Dioxide by Time-Resolved
  Harmonic Spectroscopy. \emph{Phys. Rev. Lett.} \textbf{2023}, \emph{131},
  243201\relax
\mciteBstWouldAddEndPuncttrue
\mciteSetBstMidEndSepPunct{\mcitedefaultmidpunct}
{\mcitedefaultendpunct}{\mcitedefaultseppunct}\relax
\EndOfBibitem
\bibitem[van~der Geest \latin{et~al.}(2023)van~der Geest, de~Boer, Murzyn,
  Jürgens, Ehrler, and Kraus]{Geest23a}
van~der Geest,~M.~L.; de~Boer,~J.~J.; Murzyn,~K.; Jürgens,~P.; Ehrler,~B.;
  Kraus,~P.~M. Transient High-Harmonic Spectroscopy in an Inorganic-Organic
  Lead Halide Perovskite. \emph{Journal of Physical Chemistry Letters}
  \textbf{2023}, \emph{14}, 10810--10818\relax
\mciteBstWouldAddEndPuncttrue
\mciteSetBstMidEndSepPunct{\mcitedefaultmidpunct}
{\mcitedefaultendpunct}{\mcitedefaultseppunct}\relax
\EndOfBibitem
\bibitem[Korobenko \latin{et~al.}(2021)Korobenko, Saha, Godfrey, Gertsvolf,
  Naumov, Villeneuve, Boltasseva, Shalaev, and Corkum]{Korobenko2021}
Korobenko,~A.; Saha,~S.; Godfrey,~A.~T.; Gertsvolf,~M.; Naumov,~A.~Y.;
  Villeneuve,~D.~M.; Boltasseva,~A.; Shalaev,~V.~M.; Corkum,~P.~B.
  High-harmonic generation in metallic titanium nitride. \emph{Nature
  Communications} \textbf{2021}, \emph{12}, 4981\relax
\mciteBstWouldAddEndPuncttrue
\mciteSetBstMidEndSepPunct{\mcitedefaultmidpunct}
{\mcitedefaultendpunct}{\mcitedefaultseppunct}\relax
\EndOfBibitem
\bibitem[Heissler \latin{et~al.}(2014)Heissler, Lugovoy, H{\"o}rlein,
  Waldecker, Wenz, Heigoldt, Khrennikov, Karsch, Krausz, Abel, \latin{et~al.}
  others]{heissler14a}
Heissler,~P.; Lugovoy,~E.; H{\"o}rlein,~R.; Waldecker,~L.; Wenz,~J.;
  Heigoldt,~M.; Khrennikov,~K.; Karsch,~S.; Krausz,~F.; Abel,~B.,
  \latin{et~al.}  Using the third state of matter: high harmonic generation
  from liquid targets. \emph{New Journal of Physics} \textbf{2014}, \emph{16},
  113045\relax
\mciteBstWouldAddEndPuncttrue
\mciteSetBstMidEndSepPunct{\mcitedefaultmidpunct}
{\mcitedefaultendpunct}{\mcitedefaultseppunct}\relax
\EndOfBibitem
\bibitem[Luu \latin{et~al.}(2018)Luu, Yin, Jain, Gaumnitz, Pertot, Ma, and
  W{\"o}rner]{luu18b}
Luu,~T.~T.; Yin,~Z.; Jain,~A.; Gaumnitz,~T.; Pertot,~Y.; Ma,~J.;
  W{\"o}rner,~H.~J. Extreme--ultraviolet high--harmonic generation in liquids.
  \emph{Nature communications} \textbf{2018}, \emph{9}, 3723\relax
\mciteBstWouldAddEndPuncttrue
\mciteSetBstMidEndSepPunct{\mcitedefaultmidpunct}
{\mcitedefaultendpunct}{\mcitedefaultseppunct}\relax
\EndOfBibitem
\bibitem[Mathijssen \latin{et~al.}(2023)Mathijssen, Mazzotta, Heinzerling,
  Eikema, and Witte]{mathijssen23a}
Mathijssen,~J.; Mazzotta,~Z.; Heinzerling,~A.~M.; Eikema,~K.~S.; Witte,~S.
  Material-specific high-order harmonic generation in laser-produced plasmas
  for varying plasma dynamics. \emph{Applied Physics B} \textbf{2023},
  \emph{129}, 91\relax
\mciteBstWouldAddEndPuncttrue
\mciteSetBstMidEndSepPunct{\mcitedefaultmidpunct}
{\mcitedefaultendpunct}{\mcitedefaultseppunct}\relax
\EndOfBibitem
\bibitem[Ganeev(2007)]{ganeev07a}
Ganeev,~R. High-order harmonic generation in a laser plasma: a review of recent
  achievements. \emph{Journal of Physics B: Atomic, Molecular and Optical
  Physics} \textbf{2007}, \emph{40}, R213\relax
\mciteBstWouldAddEndPuncttrue
\mciteSetBstMidEndSepPunct{\mcitedefaultmidpunct}
{\mcitedefaultendpunct}{\mcitedefaultseppunct}\relax
\EndOfBibitem
\bibitem[Dudovich \latin{et~al.}(2006)Dudovich, Smirnova, Levesque, Ivanov,
  Villeneuve, and Corkum]{dudovich06a}
Dudovich,~N.; Smirnova,~O.; Levesque,~J.; Ivanov,~M.; Villeneuve,~D.~M.;
  Corkum,~P.~B. Measuring and controlling the birth of attosecond pulses.
  \emph{Nature Physics} \textbf{2006}, \emph{2}, 781\relax
\mciteBstWouldAddEndPuncttrue
\mciteSetBstMidEndSepPunct{\mcitedefaultmidpunct}
{\mcitedefaultendpunct}{\mcitedefaultseppunct}\relax
\EndOfBibitem
\bibitem[{Roscam Abbing} \latin{et~al.}(2020){Roscam Abbing}, Campi, Sajjadian,
  Lin, Smorenburg, and Kraus]{RoscamAbbing2020}
{Roscam Abbing},~S.; Campi,~F.; Sajjadian,~F.; Lin,~N.; Smorenburg,~P.;
  Kraus,~P.~M. Divergence Control of High-Harmonic Generation. \emph{Phys. Rev.
  Appl.} \textbf{2020}, \emph{13}, 054029\relax
\mciteBstWouldAddEndPuncttrue
\mciteSetBstMidEndSepPunct{\mcitedefaultmidpunct}
{\mcitedefaultendpunct}{\mcitedefaultseppunct}\relax
\EndOfBibitem
\bibitem[Roscam~Abbing \latin{et~al.}(2021)Roscam~Abbing, Campi, Zeltsi,
  Smorenburg, and Kraus]{roscam21a}
Roscam~Abbing,~S.~D.; Campi,~F.; Zeltsi,~A.; Smorenburg,~P.; Kraus,~P.~M.
  Divergence and efficiency optimization in polarization-controlled two-color
  high-harmonic generation. \emph{Scientific Reports} \textbf{2021}, \emph{11},
  1--11\relax
\mciteBstWouldAddEndPuncttrue
\mciteSetBstMidEndSepPunct{\mcitedefaultmidpunct}
{\mcitedefaultendpunct}{\mcitedefaultseppunct}\relax
\EndOfBibitem
\bibitem[Sakai \latin{et~al.}(2003)Sakai, Minemoto, Nanjo, Tanji, and
  Suzuki]{sakai03a}
Sakai,~H.; Minemoto,~S.; Nanjo,~H.; Tanji,~H.; Suzuki,~T. Controlling the
  Orientation of Polar Molecules with Combined Electrostatic and Pulsed,
  Nonresonant Laser Fields. \emph{Phys. Rev. Lett.} \textbf{2003}, \emph{90},
  083001\relax
\mciteBstWouldAddEndPuncttrue
\mciteSetBstMidEndSepPunct{\mcitedefaultmidpunct}
{\mcitedefaultendpunct}{\mcitedefaultseppunct}\relax
\EndOfBibitem
\bibitem[Rupenyan \latin{et~al.}(2013)Rupenyan, Kraus, Schneider, and
  W\"orner]{rupenyan13a}
Rupenyan,~A.; Kraus,~P.~M.; Schneider,~J.; W\"orner,~H.~J. High-harmonic
  spectroscopy of isoelectronic molecules: Wavelength scaling of
  electronic-structure and multielectron effects. \emph{Phys. Rev. A}
  \textbf{2013}, \emph{87}, 033409\relax
\mciteBstWouldAddEndPuncttrue
\mciteSetBstMidEndSepPunct{\mcitedefaultmidpunct}
{\mcitedefaultendpunct}{\mcitedefaultseppunct}\relax
\EndOfBibitem
\bibitem[Frumker \latin{et~al.}(2012)Frumker, Hebeisen, Kajumba, Bertrand,
  W{\"o}rner, Spanner, Villeneuve, Naumov, and Corkum]{frumker12a}
Frumker,~E.; Hebeisen,~C.~T.; Kajumba,~N.; Bertrand,~J.~B.; W{\"o}rner,~H.~J.;
  Spanner,~M.; Villeneuve,~D.~M.; Naumov,~A.; Corkum,~P.~B. Oriented Rotational
  Wave-Packet Dynamics Studies via High Harmonic Generation. \emph{Physical
  Review Letters} \textbf{2012}, \emph{109}, 113901\relax
\mciteBstWouldAddEndPuncttrue
\mciteSetBstMidEndSepPunct{\mcitedefaultmidpunct}
{\mcitedefaultendpunct}{\mcitedefaultseppunct}\relax
\EndOfBibitem
\bibitem[Kraus \latin{et~al.}(2012)Kraus, Rupenyan, and W{\"o}rner]{kraus12c}
Kraus,~P.~M.; Rupenyan,~A.; W{\"o}rner,~H.~J. High-harmonic spectroscopy of
  oriented OCS molecules: emission of even and odd harmonics. \emph{Phys. Rev.
  Lett.} \textbf{2012}, \emph{109}, 233903\relax
\mciteBstWouldAddEndPuncttrue
\mciteSetBstMidEndSepPunct{\mcitedefaultmidpunct}
{\mcitedefaultendpunct}{\mcitedefaultseppunct}\relax
\EndOfBibitem
\bibitem[Kraus \latin{et~al.}(2014)Kraus, Baykusheva, and W{\"o}rner]{kraus14b}
Kraus,~P.~M.; Baykusheva,~D.; W{\"o}rner,~H.~J. Two-pulse orientation dynamics
  and high-harmonic spectroscopy of strongly oriented molecules. \emph{J. Phys.
  B} \textbf{2014}, \emph{47}, 124030\relax
\mciteBstWouldAddEndPuncttrue
\mciteSetBstMidEndSepPunct{\mcitedefaultmidpunct}
{\mcitedefaultendpunct}{\mcitedefaultseppunct}\relax
\EndOfBibitem
\bibitem[Kraus \latin{et~al.}(2015)Kraus, Tolstikhin, Baykusheva, Rupenyan,
  Schneider, Bisgaard, Morishita, Jensen, Madsen, and W\"{o}rner]{kraus15a}
Kraus,~P.~M.; Tolstikhin,~O.~I.; Baykusheva,~D.; Rupenyan,~A.; Schneider,~J.;
  Bisgaard,~C.~Z.; Morishita,~T.; Jensen,~F.; Madsen,~L.~B.; W\"{o}rner,~H.~J.
  Observation of laser-induced electronic structure in oriented polyatomic
  molecules. \emph{Nat. Commun.} \textbf{2015}, \emph{6}, 7039\relax
\mciteBstWouldAddEndPuncttrue
\mciteSetBstMidEndSepPunct{\mcitedefaultmidpunct}
{\mcitedefaultendpunct}{\mcitedefaultseppunct}\relax
\EndOfBibitem
\bibitem[W\"{o}rner \latin{et~al.}(2010)W\"{o}rner, Bertrand, Kartashov,
  Corkum, and Villeneuve]{woerner10b}
W\"{o}rner,~H.~J.; Bertrand,~J.~B.; Kartashov,~D.~V.; Corkum,~P.~B.;
  Villeneuve,~D.~M. Following a chemical reaction using high-harmonic
  interferometry. \emph{Nature} \textbf{2010}, \emph{466}, 604--607\relax
\mciteBstWouldAddEndPuncttrue
\mciteSetBstMidEndSepPunct{\mcitedefaultmidpunct}
{\mcitedefaultendpunct}{\mcitedefaultseppunct}\relax
\EndOfBibitem
\bibitem[Kraus \latin{et~al.}(2012)Kraus, Arasaki, Bertrand, Patchkovskii,
  Corkum, Villeneuve, Takatsuka, and W\"orner]{kraus12a}
Kraus,~P.~M.; Arasaki,~Y.; Bertrand,~J.~B.; Patchkovskii,~S.; Corkum,~P.~B.;
  Villeneuve,~D.~M.; Takatsuka,~K.; W\"orner,~H.~J. Time-resolved high-harmonic
  spectroscopy of nonadiabatic dynamics in NO${}_{2}$. \emph{Phys. Rev. A}
  \textbf{2012}, \emph{85}, 043409\relax
\mciteBstWouldAddEndPuncttrue
\mciteSetBstMidEndSepPunct{\mcitedefaultmidpunct}
{\mcitedefaultendpunct}{\mcitedefaultseppunct}\relax
\EndOfBibitem
\bibitem[Ruf \latin{et~al.}(2012)Ruf, Handschin, Ferre, Thire, Bertrand,
  Bonnet, Cireasa, Constant, Corkum, Descamps, Fabre, Larregaray, Mevel, Petit,
  Pons, Staedter, W\"{o}rner, Villeneuve, Mairesse, Halvick, and
  Blanchet]{ruf12a}
Ruf,~H. \latin{et~al.}  High-harmonic transient grating spectroscopy of NO$_2$
  electronic relaxation. \emph{The Journal of Chemical Physics} \textbf{2012},
  \emph{137}, 224303\relax
\mciteBstWouldAddEndPuncttrue
\mciteSetBstMidEndSepPunct{\mcitedefaultmidpunct}
{\mcitedefaultendpunct}{\mcitedefaultseppunct}\relax
\EndOfBibitem
\bibitem[Kraus and W{\"o}rner(2013)Kraus, and W{\"o}rner]{kraus12b}
Kraus,~P.~M.; W{\"o}rner,~H.~J. Time-resolved high-harmonic spectroscopy of
  valence electron dynamics. \emph{Chemical Physics} \textbf{2013}, \emph{414},
  32 -- 44\relax
\mciteBstWouldAddEndPuncttrue
\mciteSetBstMidEndSepPunct{\mcitedefaultmidpunct}
{\mcitedefaultendpunct}{\mcitedefaultseppunct}\relax
\EndOfBibitem
\bibitem[Kraus \latin{et~al.}(2013)Kraus, Zhang, Gijsbertsen, Lucchese,
  Rohringer, and W\"orner]{kraus13b}
Kraus,~P.~M.; Zhang,~S.~B.; Gijsbertsen,~A.; Lucchese,~R.~R.; Rohringer,~N.;
  W\"orner,~H.~J. High-Harmonic Probing of Electronic Coherence in Dynamically
  Aligned Molecules. \emph{Phys. Rev. Lett.} \textbf{2013}, \emph{111},
  243005\relax
\mciteBstWouldAddEndPuncttrue
\mciteSetBstMidEndSepPunct{\mcitedefaultmidpunct}
{\mcitedefaultendpunct}{\mcitedefaultseppunct}\relax
\EndOfBibitem
\bibitem[Baykusheva \latin{et~al.}(2014)Baykusheva, Kraus, Zhang, Rohringer,
  and W\"{o}rner]{baykusheva14a}
Baykusheva,~D.; Kraus,~P.~M.; Zhang,~S.~B.; Rohringer,~N.; W\"{o}rner,~H.~J.
  The sensitivities of high-harmonic generation and strong-field ionization to
  coupled electronic and nuclear dynamics. \emph{Faraday Discussions}
  \textbf{2014}, \emph{171}, 113\relax
\mciteBstWouldAddEndPuncttrue
\mciteSetBstMidEndSepPunct{\mcitedefaultmidpunct}
{\mcitedefaultendpunct}{\mcitedefaultseppunct}\relax
\EndOfBibitem
\bibitem[Zhang \latin{et~al.}(2015)Zhang, Baykusheva, Kraus, W\"orner, and
  Rohringer]{zhang15b}
Zhang,~S.~B.; Baykusheva,~D.; Kraus,~P.~M.; W\"orner,~H.~J.; Rohringer,~N.
  Theoretical study of molecular electronic and rotational coherences by
  high-order-harmonic generation. \emph{Phys. Rev. A} \textbf{2015}, \emph{91},
  023421\relax
\mciteBstWouldAddEndPuncttrue
\mciteSetBstMidEndSepPunct{\mcitedefaultmidpunct}
{\mcitedefaultendpunct}{\mcitedefaultseppunct}\relax
\EndOfBibitem
\bibitem[Hohenleutner \latin{et~al.}(2015)Hohenleutner, Langer, Schubert,
  Knorr, Huttner, Koch, Kira, and Huber]{Hohenleutner2015}
Hohenleutner,~M.; Langer,~F.; Schubert,~O.; Knorr,~M.; Huttner,~U.;
  Koch,~S.~W.; Kira,~M.; Huber,~R. Real-time observation of interfering crystal
  electrons in high-harmonic generation. \emph{Nature} \textbf{2015},
  \emph{523}, 572--575\relax
\mciteBstWouldAddEndPuncttrue
\mciteSetBstMidEndSepPunct{\mcitedefaultmidpunct}
{\mcitedefaultendpunct}{\mcitedefaultseppunct}\relax
\EndOfBibitem
\bibitem[Abbing \latin{et~al.}(2022)Abbing, Kolkowski, Zhang, Campi,
  L{\"o}tgering, Koenderink, and Kraus]{roscam22a}
Abbing,~S. D.~R.; Kolkowski,~R.; Zhang,~Z.-Y.; Campi,~F.; L{\"o}tgering,~L.;
  Koenderink,~A.~F.; Kraus,~P.~M. Extreme-ultraviolet shaping and imaging by
  high-harmonic generation from nanostructured silica. \emph{Physical review
  letters} \textbf{2022}, \emph{128}, 223902\relax
\mciteBstWouldAddEndPuncttrue
\mciteSetBstMidEndSepPunct{\mcitedefaultmidpunct}
{\mcitedefaultendpunct}{\mcitedefaultseppunct}\relax
\EndOfBibitem
\bibitem[Korobenko \latin{et~al.}(2022)Korobenko, Rashid, Heide, Naumov, Reis,
  Berini, Corkum, and Vampa]{korobenko22a}
Korobenko,~A.; Rashid,~S.; Heide,~C.; Naumov,~A.~Y.; Reis,~D.~A.; Berini,~P.;
  Corkum,~P.~B.; Vampa,~G. In-Situ Nanoscale Focusing of Extreme Ultraviolet
  Solid-State High Harmonics. \emph{Physical Review X} \textbf{2022},
  \emph{12}, 041036\relax
\mciteBstWouldAddEndPuncttrue
\mciteSetBstMidEndSepPunct{\mcitedefaultmidpunct}
{\mcitedefaultendpunct}{\mcitedefaultseppunct}\relax
\EndOfBibitem
\bibitem[Franz \latin{et~al.}(2019)Franz, Kaassamani, Gauthier, Nicolas,
  Kholodtsova, Douillard, Gomes, Lavoute, Gaponov, Ducros, \latin{et~al.}
  others]{franz19a}
Franz,~D.; Kaassamani,~S.; Gauthier,~D.; Nicolas,~R.; Kholodtsova,~M.;
  Douillard,~L.; Gomes,~J.-T.; Lavoute,~L.; Gaponov,~D.; Ducros,~N.,
  \latin{et~al.}  All semiconductor enhanced high-harmonic generation from a
  single nanostructured cone. \emph{Sci. Rep.} \textbf{2019}, \emph{9},
  5663\relax
\mciteBstWouldAddEndPuncttrue
\mciteSetBstMidEndSepPunct{\mcitedefaultmidpunct}
{\mcitedefaultendpunct}{\mcitedefaultseppunct}\relax
\EndOfBibitem
\bibitem[Murzyn \latin{et~al.}(2023)Murzyn, Guery, Nie, van~der Geest, and
  Kraus]{Murzyn2023}
Murzyn,~K.; Guery,~L.; Nie,~Z.; van~der Geest,~M.; Kraus,~P.~M. Point-spread
  function reduction through high-harmonic generation deactivation. Conference
  on Lasers and Electro-Optics/Europe (CLEO/Europe 2023) and European Quantum
  Electronics Conference (EQEC 2023). 2023; p cf\_2\_3\relax
\mciteBstWouldAddEndPuncttrue
\mciteSetBstMidEndSepPunct{\mcitedefaultmidpunct}
{\mcitedefaultendpunct}{\mcitedefaultseppunct}\relax
\EndOfBibitem
\bibitem[Nishidome \latin{et~al.}(2020)Nishidome, Nagai, Uchida, Ichinose,
  Yomogida, Miyata, Tanaka, and Yanagi]{Nishidome2020}
Nishidome,~H.; Nagai,~K.; Uchida,~K.; Ichinose,~Y.; Yomogida,~Y.; Miyata,~Y.;
  Tanaka,~K.; Yanagi,~K. Control of High-Harmonic Generation by Tuning the
  Electronic Structure and Carrier Injection. \emph{Nano Letters}
  \textbf{2020}, \emph{20}, 6215--6221\relax
\mciteBstWouldAddEndPuncttrue
\mciteSetBstMidEndSepPunct{\mcitedefaultmidpunct}
{\mcitedefaultendpunct}{\mcitedefaultseppunct}\relax
\EndOfBibitem
\bibitem[Luu and Wörner(2018)Luu, and Wörner]{Luu2018}
Luu,~T.~T.; Wörner,~H.~J. Observing broken inversion symmetry in solids using
  two-color high-order harmonic spectroscopy. \emph{Physical Review A}
  \textbf{2018}, \emph{98}, 041802(R)\relax
\mciteBstWouldAddEndPuncttrue
\mciteSetBstMidEndSepPunct{\mcitedefaultmidpunct}
{\mcitedefaultendpunct}{\mcitedefaultseppunct}\relax
\EndOfBibitem
\bibitem[Jia \latin{et~al.}(2020)Jia, Zhang, Yang, Liu, Si, Zhang, and
  Liu]{Jia2020}
Jia,~L.; Zhang,~Z.; Yang,~D.~Z.; Liu,~Y.; Si,~M.~S.; Zhang,~G.~P.; Liu,~Y.~S.
  Optical high-order harmonic generation as a structural characterization tool.
  \emph{Physical Review B} \textbf{2020}, \emph{101}, 144304\relax
\mciteBstWouldAddEndPuncttrue
\mciteSetBstMidEndSepPunct{\mcitedefaultmidpunct}
{\mcitedefaultendpunct}{\mcitedefaultseppunct}\relax
\EndOfBibitem
\bibitem[Liu \latin{et~al.}(2017)Liu, Li, You, Ghimire, Heinz, and
  Reis]{Liu2017}
Liu,~H.; Li,~Y.; You,~Y.~S.; Ghimire,~S.; Heinz,~T.~F.; Reis,~D.~A.
  High-harmonic generation from an atomically thin semiconductor. \emph{Nature
  Physics} \textbf{2017}, \emph{13}, 262--265\relax
\mciteBstWouldAddEndPuncttrue
\mciteSetBstMidEndSepPunct{\mcitedefaultmidpunct}
{\mcitedefaultendpunct}{\mcitedefaultseppunct}\relax
\EndOfBibitem
\bibitem[Ward(1965)]{Ward1965}
Ward,~J.~F. Calculation of Nonlinear Optical Susceptibilities Using
  Diagrammatic Perturbation Theory I. Introduction. \emph{Reviews of Modern
  Physics} \textbf{1965}, \emph{37}, 1\relax
\mciteBstWouldAddEndPuncttrue
\mciteSetBstMidEndSepPunct{\mcitedefaultmidpunct}
{\mcitedefaultendpunct}{\mcitedefaultseppunct}\relax
\EndOfBibitem
\bibitem[Lindberg and Koch(1998)Lindberg, and Koch]{Lindeberg1998}
Lindberg,~M.; Koch,~S.~W. Effective Bloch equations for semiconductors.
  \emph{Physical Review B} \textbf{1998}, \emph{38}, 15--1988\relax
\mciteBstWouldAddEndPuncttrue
\mciteSetBstMidEndSepPunct{\mcitedefaultmidpunct}
{\mcitedefaultendpunct}{\mcitedefaultseppunct}\relax
\EndOfBibitem
\bibitem[Haug and Koch(2004)Haug, and Koch]{Haug2004}
Haug,~H.; Koch,~S.~W. \emph{Quantum theory of the optical and electronic
  properties of semiconductors}; World Scientific, 2004; p 453\relax
\mciteBstWouldAddEndPuncttrue
\mciteSetBstMidEndSepPunct{\mcitedefaultmidpunct}
{\mcitedefaultendpunct}{\mcitedefaultseppunct}\relax
\EndOfBibitem
\bibitem[Yue and Gaarde(2022)Yue, and Gaarde]{Yue2021}
Yue,~L.; Gaarde,~M.~B. Introduction to theory of high-harmonic generation in
  solids: tutorial. \emph{Journal of the Optical Society of America B}
  \textbf{2022}, \emph{39}, 535--555\relax
\mciteBstWouldAddEndPuncttrue
\mciteSetBstMidEndSepPunct{\mcitedefaultmidpunct}
{\mcitedefaultendpunct}{\mcitedefaultseppunct}\relax
\EndOfBibitem
\bibitem[Margalit \latin{et~al.}(2021)Margalit, Lu, Çagri Top, and
  Ketterle]{Margalit}
Margalit,~Y.; Lu,~Y.-K.; Çagri Top,~F.; Ketterle,~W. Pauli blocking of light
  scattering in degenerate fermions. \emph{Science} \textbf{2021}, \emph{374},
  976--979\relax
\mciteBstWouldAddEndPuncttrue
\mciteSetBstMidEndSepPunct{\mcitedefaultmidpunct}
{\mcitedefaultendpunct}{\mcitedefaultseppunct}\relax
\EndOfBibitem
\bibitem[Chong \latin{et~al.}(2010)Chong, Min, and Xie]{Chong2010}
Chong,~S.; Min,~W.; Xie,~X.~S. Ground-state depletion microscopy: Detection
  sensitivity of single-molecule optical absorption at room temperature.
  \emph{Journal of Physical Chemistry Letters} \textbf{2010}, \emph{1},
  3316--3322\relax
\mciteBstWouldAddEndPuncttrue
\mciteSetBstMidEndSepPunct{\mcitedefaultmidpunct}
{\mcitedefaultendpunct}{\mcitedefaultseppunct}\relax
\EndOfBibitem
\bibitem[Jannin \latin{et~al.}(2022)Jannin, van~der Werf, Steinebach, Bethlem,
  and Eikema]{Jannin2022}
Jannin,~R.; van~der Werf,~Y.; Steinebach,~K.; Bethlem,~H.~L.; Eikema,~K.~S.
  Pauli blocking of stimulated emission in a degenerate Fermi gas. \emph{Nature
  Communications} \textbf{2022}, \emph{13}, 6479\relax
\mciteBstWouldAddEndPuncttrue
\mciteSetBstMidEndSepPunct{\mcitedefaultmidpunct}
{\mcitedefaultendpunct}{\mcitedefaultseppunct}\relax
\EndOfBibitem
\bibitem[Kastner(1992)]{Kastner1992}
Kastner,~M.~A. The single-electron transistor. \emph{Review of Modern Physics}
  \textbf{1992}, \emph{64}, 849\relax
\mciteBstWouldAddEndPuncttrue
\mciteSetBstMidEndSepPunct{\mcitedefaultmidpunct}
{\mcitedefaultendpunct}{\mcitedefaultseppunct}\relax
\EndOfBibitem
\bibitem[Vampa \latin{et~al.}(2014)Vampa, McDonald, Orlando, Klug, Corkum, and
  Brabec]{Vampa2014}
Vampa,~G.; McDonald,~C.~R.; Orlando,~G.; Klug,~D.~D.; Corkum,~P.~B.; Brabec,~T.
  Theoretical analysis of high-harmonic generation in solids. \emph{Physical
  Review Letters} \textbf{2014}, \emph{113}, 073901\relax
\mciteBstWouldAddEndPuncttrue
\mciteSetBstMidEndSepPunct{\mcitedefaultmidpunct}
{\mcitedefaultendpunct}{\mcitedefaultseppunct}\relax
\EndOfBibitem
\bibitem[Becker \latin{et~al.}(1988)Becker, Fragnito, Cruz, Fork, Cunningham,
  Henry, and Shank]{Becker1988}
Becker,~P.~C.; Fragnito,~H.~L.; Cruz,~C. H.~B.; Fork,~R.~L.; Cunningham,~J.~E.;
  Henry,~J.~E.; Shank,~C.~U. Femtosecond Photon Echoes from Band-to-Band
  Transitions in GaAs. \emph{Physical Review Letters} \textbf{1988}, \emph{61},
  1647\relax
\mciteBstWouldAddEndPuncttrue
\mciteSetBstMidEndSepPunct{\mcitedefaultmidpunct}
{\mcitedefaultendpunct}{\mcitedefaultseppunct}\relax
\EndOfBibitem
\bibitem[Kilen \latin{et~al.}(2020)Kilen, Kolesik, Hader, Moloney, Huttner,
  Hagen, and Koch]{Kilen2020}
Kilen,~I.; Kolesik,~M.; Hader,~J.; Moloney,~J.~V.; Huttner,~U.; Hagen,~M.~K.;
  Koch,~S.~W. Propagation Induced Dephasing in Semiconductor High-Harmonic
  Generation. \emph{Physical Review Letters} \textbf{2020}, \emph{125},
  083901\relax
\mciteBstWouldAddEndPuncttrue
\mciteSetBstMidEndSepPunct{\mcitedefaultmidpunct}
{\mcitedefaultendpunct}{\mcitedefaultseppunct}\relax
\EndOfBibitem
\bibitem[Brown \latin{et~al.}(2022)Brown, Álvaro Jiménez-Galán, Silva, and
  Ivanov]{Brown2022}
Brown,~G.~G.; Álvaro Jiménez-Galán,; Silva,~R. E.~F.; Ivanov,~M. A
  Real-Space Perspective on Dephasing in Solid-State High Harmonic Generation.
  \emph{arXiv:2210.16889} \textbf{2022}, \relax
\mciteBstWouldAddEndPunctfalse
\mciteSetBstMidEndSepPunct{\mcitedefaultmidpunct}
{}{\mcitedefaultseppunct}\relax
\EndOfBibitem
\bibitem[Vampa \latin{et~al.}(2014)Vampa, McDonald, Orlando, Klug, Corkum, and
  Brabec]{vampa14a}
Vampa,~G.; McDonald,~C.; Orlando,~G.; Klug,~D.; Corkum,~P.; Brabec,~T.
  Theoretical analysis of high-harmonic generation in solids. \emph{Physical
  review letters} \textbf{2014}, \emph{113}, 073901\relax
\mciteBstWouldAddEndPuncttrue
\mciteSetBstMidEndSepPunct{\mcitedefaultmidpunct}
{\mcitedefaultendpunct}{\mcitedefaultseppunct}\relax
\EndOfBibitem
\bibitem[Tancogne-Dejean \latin{et~al.}(2017)Tancogne-Dejean, M{\"u}cke,
  K{\"a}rtner, and Rubio]{tancogne17a}
Tancogne-Dejean,~N.; M{\"u}cke,~O.~D.; K{\"a}rtner,~F.~X.; Rubio,~A.
  Ellipticity dependence of high-harmonic generation in solids originating from
  coupled intraband and interband dynamics. \emph{Nature communications}
  \textbf{2017}, \emph{8}, 745\relax
\mciteBstWouldAddEndPuncttrue
\mciteSetBstMidEndSepPunct{\mcitedefaultmidpunct}
{\mcitedefaultendpunct}{\mcitedefaultseppunct}\relax
\EndOfBibitem
\bibitem[Bennett \latin{et~al.}(1990)Bennett, Soref, and Del~Alamo]{bennett90a}
Bennett,~B.~R.; Soref,~R.~A.; Del~Alamo,~J.~A. Carrier-induced change in
  refractive index of InP, GaAs and InGaAsP. \emph{IEEE Journal of Quantum
  Electronics} \textbf{1990}, \emph{26}, 113--122\relax
\mciteBstWouldAddEndPuncttrue
\mciteSetBstMidEndSepPunct{\mcitedefaultmidpunct}
{\mcitedefaultendpunct}{\mcitedefaultseppunct}\relax
\EndOfBibitem
\bibitem[Imada \latin{et~al.}(1998)Imada, Fujimori, and Tokura]{imada1998metal}
Imada,~M.; Fujimori,~A.; Tokura,~Y. Metal-insulator transitions. \emph{Reviews
  of modern physics} \textbf{1998}, \emph{70}, 1039\relax
\mciteBstWouldAddEndPuncttrue
\mciteSetBstMidEndSepPunct{\mcitedefaultmidpunct}
{\mcitedefaultendpunct}{\mcitedefaultseppunct}\relax
\EndOfBibitem
\bibitem[Wegkamp and St{\"a}hler(2015)Wegkamp, and
  St{\"a}hler]{wegkamp2015ultrafast}
Wegkamp,~D.; St{\"a}hler,~J. Ultrafast dynamics during the photoinduced phase
  transition in VO2. \emph{Progress in Surface Science} \textbf{2015},
  \emph{90}, 464--502\relax
\mciteBstWouldAddEndPuncttrue
\mciteSetBstMidEndSepPunct{\mcitedefaultmidpunct}
{\mcitedefaultendpunct}{\mcitedefaultseppunct}\relax
\EndOfBibitem
\bibitem[Wegkamp \latin{et~al.}(2014)Wegkamp, Herzog, Xian, Gatti, Cudazzo,
  McGahan, Marvel, Haglund~Jr, Rubio, Wolf, \latin{et~al.}
  others]{wegkamp2014instantaneous}
Wegkamp,~D.; Herzog,~M.; Xian,~L.; Gatti,~M.; Cudazzo,~P.; McGahan,~C.~L.;
  Marvel,~R.~E.; Haglund~Jr,~R.~F.; Rubio,~A.; Wolf,~M., \latin{et~al.}
  Instantaneous band gap collapse in photoexcited monoclinic VO 2 due to
  photocarrier doping. \emph{Physical review letters} \textbf{2014},
  \emph{113}, 216401\relax
\mciteBstWouldAddEndPuncttrue
\mciteSetBstMidEndSepPunct{\mcitedefaultmidpunct}
{\mcitedefaultendpunct}{\mcitedefaultseppunct}\relax
\EndOfBibitem
\bibitem[Wang \latin{et~al.}(2022)Wang, Nie, Shi, Wang, and
  Wang]{wang2022coherent}
Wang,~Y.; Nie,~Z.; Shi,~Y.; Wang,~Y.; Wang,~F. Coherent vibrational dynamics of
  NbO 2 film. \emph{Physical Review Materials} \textbf{2022}, \emph{6},
  035005\relax
\mciteBstWouldAddEndPuncttrue
\mciteSetBstMidEndSepPunct{\mcitedefaultmidpunct}
{\mcitedefaultendpunct}{\mcitedefaultseppunct}\relax
\EndOfBibitem
\bibitem[Yang \latin{et~al.}(2011)Yang, Ko, and Ramanathan]{yang2011oxide}
Yang,~Z.; Ko,~C.; Ramanathan,~S. Oxide electronics utilizing ultrafast
  metal-insulator transitions. \emph{Annual Review of Materials Research}
  \textbf{2011}, \emph{41}, 337--367\relax
\mciteBstWouldAddEndPuncttrue
\mciteSetBstMidEndSepPunct{\mcitedefaultmidpunct}
{\mcitedefaultendpunct}{\mcitedefaultseppunct}\relax
\EndOfBibitem
\bibitem[Kumar \latin{et~al.}(2020)Kumar, Williams, and Wang]{kumar2020third}
Kumar,~S.; Williams,~R.~S.; Wang,~Z. Third-order nanocircuit elements for
  neuromorphic engineering. \emph{Nature} \textbf{2020}, \emph{585},
  518--523\relax
\mciteBstWouldAddEndPuncttrue
\mciteSetBstMidEndSepPunct{\mcitedefaultmidpunct}
{\mcitedefaultendpunct}{\mcitedefaultseppunct}\relax
\EndOfBibitem
\bibitem[Li \latin{et~al.}(2019)Li, Fan, Ma, Wang, and Jin]{Li2019}
Li,~X.; Fan,~J.; Ma,~J.; Wang,~G.; Jin,~C. Application of optimized waveforms
  for enhancing high-harmonic yields in a three-color laser-field synthesizer.
  \emph{Optics Express} \textbf{2019}, \emph{27}, 841\relax
\mciteBstWouldAddEndPuncttrue
\mciteSetBstMidEndSepPunct{\mcitedefaultmidpunct}
{\mcitedefaultendpunct}{\mcitedefaultseppunct}\relax
\EndOfBibitem
\bibitem[Kroh \latin{et~al.}(2018)Kroh, Jin, Krogen, Keathley, Calendron,
  Siqueira, Liang, Falcão-Filho, Lin, Kärtner, and Hong]{Kroh2018}
Kroh,~T.; Jin,~C.; Krogen,~P.; Keathley,~P.~D.; Calendron,~A.-L.;
  Siqueira,~J.~P.; Liang,~H.; Falcão-Filho,~E.~L.; Lin,~C.~D.;
  Kärtner,~F.~X.; Hong,~K.-H. Enhanced high-harmonic generation up to the soft
  X-ray region driven by mid-infrared pulses mixed with their third harmonic.
  \emph{Optics Express} \textbf{2018}, \emph{26}, 16955\relax
\mciteBstWouldAddEndPuncttrue
\mciteSetBstMidEndSepPunct{\mcitedefaultmidpunct}
{\mcitedefaultendpunct}{\mcitedefaultseppunct}\relax
\EndOfBibitem
\bibitem[Liu \latin{et~al.}(2016)Liu, Zheng, Zeng, and Li]{Liu2016}
Liu,~C.; Zheng,~Y.; Zeng,~Z.; Li,~R. Effect of elliptical polarization of
  driving field on high-order-harmonic generation in semiconductor ZnO.
  \emph{Physical Review A} \textbf{2016}, \emph{93}, 043806\relax
\mciteBstWouldAddEndPuncttrue
\mciteSetBstMidEndSepPunct{\mcitedefaultmidpunct}
{\mcitedefaultendpunct}{\mcitedefaultseppunct}\relax
\EndOfBibitem
\bibitem[Hollinger \latin{et~al.}(2021)Hollinger, Herrmann, Korolev, Zapf,
  Shumakova, Röder, Uschmann, Pugžlys, Baltuška, Zürch, Ronning, Spielmann,
  and Kartashov]{Hollinger2021}
Hollinger,~R.; Herrmann,~P.; Korolev,~V.; Zapf,~M.; Shumakova,~V.; Röder,~R.;
  Uschmann,~I.; Pugžlys,~A.; Baltuška,~A.; Zürch,~M.; Ronning,~C.;
  Spielmann,~C.; Kartashov,~D. Polarization dependent excitation and high
  harmonic generation from intense mid-IR laser pulses in ZnO.
  \emph{Nanomaterials} \textbf{2021}, \emph{11}, 1--10\relax
\mciteBstWouldAddEndPuncttrue
\mciteSetBstMidEndSepPunct{\mcitedefaultmidpunct}
{\mcitedefaultendpunct}{\mcitedefaultseppunct}\relax
\EndOfBibitem
\bibitem[Rodnyi and Khodyuk(2011)Rodnyi, and Khodyuk]{Rodnyi2011}
Rodnyi,~P.~A.; Khodyuk,~I.~V. Optical and Luminescence Properties of Zinc
  Oxide. \emph{Optics and spectroscopy} \textbf{2011}, \emph{111},
  776--785\relax
\mciteBstWouldAddEndPuncttrue
\mciteSetBstMidEndSepPunct{\mcitedefaultmidpunct}
{\mcitedefaultendpunct}{\mcitedefaultseppunct}\relax
\EndOfBibitem
\bibitem[Kivshar(2018)]{2018Kivshar}
Kivshar,~Y. {All-dielectric meta-optics and non-linear nanophotonics}.
  \emph{National Science Review} \textbf{2018}, \emph{5}, 144--158\relax
\mciteBstWouldAddEndPuncttrue
\mciteSetBstMidEndSepPunct{\mcitedefaultmidpunct}
{\mcitedefaultendpunct}{\mcitedefaultseppunct}\relax
\EndOfBibitem
\end{mcitethebibliography}

\end{document}